\documentclass[%
 aip,
 rsi,
 amsmath,amssymb,
 reprint,
]{revtex4-1}

\usepackage{graphicx}
\usepackage{dcolumn}
\usepackage{bm}
\usepackage[colorlinks=true,linkcolor=blue,citecolor=cyan]{hyperref}
\usepackage[version=4]{mhchem}
\usepackage{float}

\usepackage{color}
\usepackage[dvipsnames]{xcolor}
\usepackage{multirow}
\usepackage{rotating}
\usepackage{threeparttable}
\usepackage{braket}
\usepackage{bbold}
\usepackage{inputenc}
\usepackage{caption}
\usepackage{subcaption}

\usepackage{xr}
\usepackage{soul}
\begin{document}


\title{Simulating core electron binding energies of halogenated species adsorbed on ice surfaces and in solution with relativistic quantum embedding calculations}

\author{Richard A. Opoku}
\author{Céline Toubin}

\author{André Severo Pereira Gomes}
\email{andre.gomes@univ-lille.fr}
\affiliation{Univ.\ Lille, CNRS, 
UMR 8523 -- PhLAM -- Physique des Lasers Atomes et Mol\'ecules, F-59000 Lille, France}

\date{\today}

\begin{abstract}
In this work we investigate the effects of the environment on the X-ray photoelectron spectra of hydrogen chloride and the chloride ions adsorbed on ice surfaces, as well as of chloride ions in water droplets. In our approach, we combine a density functional theory (DFT) description of the ice surface with that of the halogen species with the recently developed relativistic core-valence separation equation of motion coupled cluster (CVS-EOM-IP-CCSD) via the frozen density embedding formalism (FDE), to determine the K and L$_{1,2,3}$ edges of chlorine. Our calculations, which incorporate temperature effects through snapshots from classical molecular dynamics simulations, are shown to reproduce experimental trends in the change of the core binding energies for Cl$^-$ upon moving from a liquid (water droplets) to an interfacial (ice quasi-liquid layer) environment.
Our simulations yield water valence band binding energies in good agreement with experiment, and that vary little between the droplets and the ice surface. For the halide core binding energies there is an overall trend of overestimating experimental values, though good agreement between theory and experiment is found for Cl$^-$ in water droplets and on ice.
For HCl on the other hand there are significant discrepancies between experimental and calculated core binding energies when we consider structural models which maintain the H-Cl bond more or less intact. An analysis of models that allow for pre-dissociated and dissociated structures suggests that experimentally observed chemical shifts in binding energies between Cl$^-$ snd HCl would reqire that H$^+$ (in the form of H$_3$O$^+$) and Cl$^-$ are separated by roughly 4-6\AA.
\end{abstract}

\maketitle

\section{Introduction\label{sec:level1}}

Ice is everywhere in the environment and its peculiar structure and properties make it a subject of intense scientific research. Studies connected to ice indicate that it hosts reactions that can influence climate, air quality, biology systems and initiate ozone destruction \cite{solomon1999stratospheric,abbatt2003interactions, kahan2010benzene}. Hydrogen Bonding (HB) between ice and trace gases is the first step towards their interaction \cite{parent2011hcl}. Investigations into the bound state of strong acids \cite{kvrepelova2010surface} interaction with ice indicate that strong acids lead to modification in the HB network of the liquid-like layer on ice surface. It has been shown however that, weak acids adsorption on ice surface does not produce any significant changes in the HB network of water ice. \cite{newberg2015adsorption}

In this respect, the influence of strong acids, in particular, hydrogen chloride (HCl) and its dissociated ionic chloride by ice has attracted a lot of attention over the years due to their link to ozone depletion~\cite{Abbatt2012}.

A better understanding of the chemical processes associated with how the reactions of these reservoir gases at the ice surface differ from that of the bulk ice surface is essential to their interpretation, and consequently to atmospheric science and environmental chemistry. Surface-specific spectroscopic approaches have been instrumental in gathering 
detailed information on structural and electronic properties of solvated halide/halide ions~\cite{smith2007effects,perera2009perturbations}, and among these X-ray Photoelectron Spectroscopy (XPS) stands out as a particularly powerful technique~\cite{hufner2013photoelectron} due to its high specificity, and the great sensitivity of core binding energies (BEs) to small perturbations to the surroundings of the atoms of interest \cite{tissot2015cation,gladich2020surface,partanen2013solvation,eriksson2014solvent,kim2013solvent}.

The surface sensitivity and chemical selectivity of radiations from XPS has made it possible to investigate the loss of gas-phase molecules as well as their behavior and transformation in complex reactions or solvent mediums. \cite{yamamoto2010water,newberg2011formation, dau2012observation,krepelova2013adsorption,tissot2015cation,olivieri2016quantitative,gaiduk2016photoelectron,starr2017combined}

This is illustrated in recent investigations of the electronic structure of halogen-related systems interacting with solvent environment \cite{parent2011hcl, eriksson2014solvent, jena2015solvent}, for which the evidence from XPS suggests the dependence of chemical and solvent binding energy induced shifts on the HB network configuration  of the solvent to the halide systems. In a pioneering work by~\citeauthor{mcneill2006hydrogen}\cite{mcneill2006hydrogen} it has been shown that the quasiliquid layer (QLL) plays an essential role in influencing the sorption behavior of HCl and on the chemistry of environmental ice surfaces, \cite{mcneill2006hydrogen, mcneill2007interaction} for which the evidence of XPS suggests that the dissociated form of HCl perturbs the HB network of the liquid-like layer on ice.~\cite{parent2011hcl, kong2017coexistence} From these studies, it is observed that the ioinization of HCl on ice surfaces follow a janus-like behavior, where molecular HCl is formed on the ice surface and its dissociated form is observed at the uppermost bulk layer of the ice. In addition, there is a long standing debate on whether the dissociation of HCl on ice surfaces is temperature dependent. While some studies show that the dissociation occurs only at high temperatures \cite{bolton2003qm, park2005adsorption}, other studies indicate that this mechanism can occur even at low temperatures. \cite{parent2011hcl, parent2005adsorption} 

As the physical and chemical processes at play (with respect to the interaction between adsorbed species and the ice surface, as well as the interaction between the incoming X-rays and the sample) are quite complex, it is difficult to make sense of the experimental results without the help of theoretical models, both in terms of geometric (the arrangement of the atoms) and electronic structure components. 

From the electronic structure standpoint the problem consists of determining core binding energy for a particular atom and edge while incorporating the effects of the environment--which may go well beyond the immediate surroundings of the atoms of interest, and may be quite severely affected by the structural changes mentioned above. That requires first the treatment of electron correlation and relaxation effects, for which the equation of motion coupled cluster (EOMCC)~\cite{1993_stanton_EOMCC,2012_Sneskov_wires_EOMRev,2012_bartlett_wires_CC_EOMCC,2008_krylov_annrevphyschem_eom,1998_christiansen_ijqc,1990_koch_jcp,EOMRESPONSE}, combined with the physically motivated CVS approximation~\cite{1980_Cederbaum_PRA_CVS}, has proven to reliably target core states in an efficient manner with little modification to standard diagonalization approaches~\cite{2015_coriani_jcp,CVS:erratum,2017_sadybekov_jcp_CoreIE,2019_vidal_jctc,halbert2020}. Second, it is essential to use relativistic Hamiltonians~\cite{saue2011relativistic} in order to capture the changes in core BEs due to scalar relativistic and spin-orbit coupling effects (the latter being responsible for the splitting of the L, M and N edges). There, methods based on the transformation from 4- to 2-component approaches are particularly useful for correlated calculations such as those with CC approaches\cite{2009_sikkema_jcp,zheng2019performance, lee2019excited}. 

Due to the relativistic correlated electronic structure methods' steep computational scaling with system size $N$ ($O(N^6)$ for CCSD-based methods), the incorporation of the environment surrounding the species of interest on the calculations is in general not possible beyond a few nearest neighbors. In this case embedding theories~\cite{gomes2012quantum,jacob2014subsystem,wesolowski2008embedding, libisch2014embedded, wesolowski2015frozen,sun2016quantum}, in which a system is partitioned into a collection of interacting subsystems, are a very cost-effective approach. Among the embedding approaches, classical embedding (QM/MM) models (continuum models, point-charge embedding, classical force fields etc) are computationally very efficient but at the cost of foregoing any prospect of extracting electronic information from the environment, and will be bound by the limitations of the classical models (e.g. the difficulties of continuum models to account for specific interactions such as hydrogen bonding). Purely quantum embedding approaches (QM/QM), on the other hand, may be more costly but with the advantage of permitting one to extract information from the electron density or wavefunctions of the environment and as such have been used to study absorption~\cite{chulhai2018projection} and reaction energies \cite{hegely2018dual}, electronically excited~\cite{gomes2008calculation, bennie2017pushing, prager2017implementation, hofener2016wave, coughtrie2018embedded, hofener2013solvatochromic, daday2014wavefunction} and ionized \cite{bouchafra2018predictive, sadybekov2017coupled} states of species of experimental interest. 

The  frozen density embedding (FDE) approach is a particularly interesting QM/QM method since it provides a framework to seamlessly combine CC and DFT approaches (CC-in-DFT) for both ground and excited states~\cite{gomes2012quantum,hofener2013solvatochromic,gomes2008calculation,hofener2012molecular}. It has been shown to successfully tackle the calculation of valence electron BEs of halogens in water droplets~\cite{bouchafra2018predictive}, while providing rather accurate valence water binding energies with no additional effort to represent the system of interest. Such an approach would be particularly interesting for addressing XPS spectra, as it would allow the calculation of the core spectra with correlated approaches for the species of interest while providing information on the valence band of the environment, which can then serve as an internal reference and help in comparisons to experiment.

To the best of our knowledge, there have been rather few studies employing QM/QM embedding for simulating core spectra:~\citeauthor{parravicini2021}~\cite{parravicini2021} have investigated projection-based CC-in-DFT embedding for the calculation of carbon K edge ionization energies (as well as valence excitation, ionization energies and resonances for first- and second-row model systems), employing non-relativistic Hamiltonians.  

Thus, in this contribution, we aim to provide a computational protocol based on relativistic CC-in-DFT calculations, that can be used in a black-box manner to obtain absolute core binding energies (which are heavily dependent on the Hamiltonan and electronic structure approach employed) for species containing atoms beyond the first row, and in complex environments. To this end, we combine the basic ingredients of the computational protocol of~\citeauthor{bouchafra2018predictive}\cite{bouchafra2018predictive} with the relativistic CVS-EOM-IP-CCSD method~\cite{halbert2020}, and investigate the performance of the resulting CVS-EOM-IP-CCSD-in-DFT method in the determination of chlorine (Cl) core electron BEs, and associated chemical shifts, for hydrogen chloride (HCl) and ionic chloride (Cl$^{-}$) adsorption on ice surfaces. We shall also profit of this investigation and determine the ionic chloride BEs on a water droplet model~\cite{bouchafra2018predictive}.

The manuscript is organized as follows: the basic theoretical aspects of DFT-in-DFT and CC-in-DFT embedding are outlined in section~\ref{sec:methods}, with a description of the structural models (alongside the details of the calculations) provided in section~\ref{sec:compdet}. The discussion of our results, and conclusions are presented in sections~\ref{sec:discussion} and~\ref{sec:conclusion}, respectively.

\section{Theoretical approaches\label{sec:methods}}
\subsection{\label{sec:level5}Frozen density embedding (FDE) method}
The main idea of FDE \cite{wesolowski1993frozen,gomes2012quantum} is the separation of the total electron density $\rho_{tot}$ of a system into a number of density subsystems. Two subsystems are considered in our case and the whole system is represented as the sum of the density of the subsystem of interest, $\rho_{\alpha}$ and subsystem of the environment, $\rho_{\beta}$ (i.e. $\rho_{tot}$ = $\rho_{\alpha}$ + $\rho_{\beta}$). The $\rho_{\beta}$ is considered to be frozen in this approximation. The corresponding total energy of the whole system is based on the electron densities of the subsystems and can then be written as
\begin{equation}
E_{tot}[\rho_{tot}] = E_{\alpha}[\rho_{\alpha}] + E_{\beta}[\rho_{\beta}] + E_{int}[\rho_{\alpha}, \rho_{\beta}] 
\label{ENE}
\end{equation}
where $E_{int}[\rho_{\alpha}, \rho_{\beta}]$ is the energy obtained from the interaction of the two subsystems, which is known as the interaction energy of the system. The interaction energy is given as
\begin{equation}
\begin{split}
E_{int}[\rho_{\alpha}, \rho_{\beta}] = E_{int}^{NN} + \int\rho_{\alpha}(r)v_{\beta}^{nuc}(r) + \rho_{\beta}(r)v_{\alpha}^{nuc}(r) \\
+ \int\frac{\rho_{\alpha}(r)\rho_{\beta}(r^{\prime})}{|r - r^{\prime}|}d^{3}rd^{3}r^{\prime}
+ E^{nadd}_{xc}[\rho_{\alpha},\rho_{\beta}] 
 + T^{nadd}_{s}[\rho_{\alpha},\rho_{\beta}] 
\label{int}
\end{split}
\end{equation}
where $E_{int}^{NN}$ is the nuclear repulsion energy between subsystems, $v_{\alpha}^{nuc}$ and $v_{\beta}^{nuc}$ are the electrostatic potential of the nuclei in subsystems $\alpha$ and $\beta$ respectively. $E_{xc}^{nadd}$ and $T_{s}^{nadd}$ are the non-additive contributions due to exchange-correlation and kinetic energies respectively are defined as 
\begin{equation}
\begin{split}
E_{xc}^{nadd}[\rho_{\alpha}, \rho_{\beta}] = E_{xc}[\rho_{\alpha} + \rho_{\beta}] - E_{xc}[\rho_{\alpha}] - E_{xc}[\rho_{\beta}]\\
T_{s}^{nadd}[\rho_{\alpha}, \rho_{\beta}] = T_{s}[\rho_{\alpha} + \rho_{\beta}] - T_{s}[\rho_{\alpha}] - T_{s}[\rho_{\beta}]
\label{ETnadd}
\end{split}
\end{equation}
The non-additive kinetic energy and the non-additive exchange-correlation energy take into consideration non-classical contributions to the energy. The  non-additive kinetic energy prevents spurious delocalization among the subsystems as observed by balancing the attractive interaction in the nuclear framework of one subsystem and the electron density of another subsystem. The FDE uses only the electron density in the calculation of interaction between subsystems without the sharing of orbital information among the subsystems. Minimization of the total energy of the system with respect to $\rho_{\alpha}$ yields an Euler-Lagrangian equation that keeps the number of electrons in the subsystem of interest fixed.\cite{gomes2012quantum}

The application of the Euler-Lagrangian equation in the FDE allows the molecular system to be subdivided into smaller interacting fragments and each of them being treated at the most suitable level of theory. Although based on DFT, the FDE scheme also allows treatment of one of the subsystems with wave function method and the rest of the subsystems with DFT (WFT-in-DFT) \cite{dresselhaus2015self, prager2017implementation} or treating all the subsystems with wave function (WFT-in-WFT) \cite{hofener2016wave}. Several literatures have implemented such WFT-in-DFT, in particular coupling of CC with DFT to accurately probe the excitation energies  \cite{hofener2013solvatochromic, gomes2012quantum, daday2014wavefunction} and ionization energies \cite{bouchafra2018predictive} of numerous molecules. 

To obtain the electron density of the subsystem of interest in Kohn-Sham (KS) DFT, the total energy, $E_{tot}[\rho_{tot}]$ is minimized concerning $\rho_{\alpha}$, while the electron density of the subsystem of the environment is kept frozen. It is performed under the restriction that the number of electrons in subsystem $\alpha$ is fixed, with the orbitals of the embedded system generated from a set of KS-like equations,
\begin{equation}
[T_{s}(i)+ v_{eff}^{KS}[\rho_{\alpha}] + v_{int}^{\alpha} [\rho_{\alpha},\rho_{\beta}] - \epsilon_{i}]\phi_{i}^{\alpha}(r)=0
\label{ksorbital}
\end{equation}
where $T_{s}(i)$ and $v^{KS}_{eff}[\rho_{\alpha}]$ are the KS kinetic energy and effective potential of the isolated subsystem of interest respectively.  The embedding potential which describes the interaction between subsystem $\alpha$ and the frozen subsystem $\beta$ is
\begin{equation}\label{ns30-emb}
\begin{split}
v^{\alpha}_{int}[\rho_{\alpha}, \rho_{\beta}](r) =  V_{\beta}^{nuc}(r) + \int \frac{\rho_{\beta}(r^{\prime})}{|r-
r^{\prime}|}dr^{\prime} \\ 
+ \frac{\delta E_{xc}[\rho]}{\delta\rho(r)}\bigg|_{\rho=\rho_{tot}} -\frac{\delta E_{xc}[\rho]}{\delta\rho(r)}\bigg|_{\rho=\rho_{\alpha}} \\
+ \frac{\delta T_{s}[\rho]}{\delta\rho(r)}\bigg|_{\rho=\rho_{tot}} + \frac{\delta T_{s}[\rho]}{\delta\rho (r)}\bigg|_{\rho=\rho_{\alpha}}
\end{split}
\end{equation}

\subsection{\label{sec:level4}Core-Valence Separation (CVS) Equation-of-motion coupled cluster (CVS-EOM-CC) theory}
In the CVS-EOM-IP-CCSD method, the BEs are obtained from
the solution of the projected eigenvector and its corresponding eigenvalue equation  \cite{2015_coriani_jcp,CVS:erratum,halbert2020}

\begin{equation}
P_{c}^{v}( \, \bar{H}P_{c}^{v}R_{k}^{IP}) \, = \Delta E_{k} P_{c}^{v}R_{k}^{IP}
\label{EOM1}
\end{equation}
where $\Delta E_{k}$ is the ionization energy of the system, $\bar{H}$ = $e^{-\hat{T}}\hat{H}e^{\hat{T}}$ is a similarity transformed Hamiltonian including equation ~\ref{ns30-emb}, $P_{c}^{v}$ is a projector  introduced to restrict all elements of valence orbitals to zero and $R_{k}^{IP}$ is the operator that transforms the coupled-cluster ground-state to electron detachment states. $R_{k}^{IP}$  is given as
\begin{equation}
R_{k}^{IP} = \sum_{i}r_{i}\{a_{i} \} + \sum_{i>j,a}r_{ij}^{a}\{a_{a}^{\dagger}a_{j}a_{i}\}
\label{EOM-ip}
\end{equation}

\section{Computational details\label{sec:compdet}}

\subsection{Electronic structure calculations}

All DFT~\cite{guerra1998towards} and DFT-in-DFT~\cite{jacob2008flexible} calculations have been performed with the 2017 version of the ADF code\cite{ADF2017}, employing the scalar relativistic zeroth-order regular approximation (ZORA) Hamiltonian\cite{lenthe1993relativistic} and triple zeta basis sets with polarization function (TZP)\cite{van2003optimized}. In the case of single-point calculations and in the determination of embedding potentials, the statistical average of orbital potentials (SAOP) model potential~\cite{grit1999,schipper2000molecular} was used for the exchange-correlation potential of the subsystems, whereas the PBE\cite{perdew1996generalized} and PW91k\cite{lembarki1994obtaining} density functionals were employed for the non-additive exchange-correlation and kinetic energy contributions, respectively. In embedding calculations no frozen cores were employed. In the case of geometry optimizations, the PBE functional was used, along with the large core option. All integration grids were taken as the default in ADF. Embedding calculations have been performed via the \textsc{PyADF} scripting framework~\cite{pyadf-2011}.

All coupled-cluster calculations were carried out with the \textsc{Dirac} electronic structure code~\cite{saue2020} (with the DIRAC19~\cite{DIRAC19} release and revisions \texttt{dbbfa6a, 0757608, 323ab67, 2628039, 1e798e5, b9f45bd}). The Dyall basis sets\cite{2002_dyall_TCA_basis,2003_dyall_TCA_basis,2016_dyall_TCA_basis} of triple-zeta quality, complemented with two diffuse functions for each angular momenta as in~\cite{bouchafra2018predictive} (d-aug-dyall.acv3z) were employed for chloride, while the Dunning aug-cc-pVTZ basis sets~\cite{1994_woon_jcp_Dunning} have been employed for hydrogen and oxygen. The basis sets were kept uncontracted in all calculations. In order to estimate the complete basis set limit (CBS) of CVS-EOM-IP-CCSD calculations, we also carried out calculations with quadruple-zeta quality bases (d-aug-dyall.acv4z and aug-cc-pVQZ) for selected systems, and used a two-point formula as carried out by~\citeauthor{bouchafra2018predictive}~\cite{bouchafra2018predictive}. 

Apart from the Dirac--Coulomb ($^4$DC) Hamiltonian, we employed the molecular mean-field\cite{2009_sikkema_jcp} approximation to the Dirac--Coulomb--Gaunt ($^2$DCG$^M$) Hamiltonian. In it the Gaunt-type integrals are explicitly taken into account only during the 4-component SCF step, as the transformation of these to MO basis is not implemented. Unless otherwise noted, we employed the usual approximation of the energy contribution from $\left(SS|SS \right )$-type two-electron integrals by a point-charge model.~\cite{1997_visscher_TCA_ssss_int_approx}. In CC-in-DFT calculations, the embedding potential obtained (with ADF) at DFT-in-DFT level is included in \textsc{Dirac} as an additional one-body operator to the Hamiltonian, following the setup outlined in~\cite{gomes2008calculation}. 

Unless otherwise noted, all occupied and virtual spinors were considered in the correlation treatment.  The core binding energy calculations with CVS-EOM-IP-CCSD~\cite{halbert2020} were performed for the K, L$_1$, L$_2$ and L$_3$ edges of the chlorine atom. The energies so obtained represent electronic states with main contributions arising from holes in the $1s, 2s, 2p_{1/2}$ and $2p_{3/2}$ spinors, respectively.

The datasets associated with this study are available at the Zenodo repository~\cite{opoku2021dataset}.

\subsection{MD-derived structures}

The structures for Cl$^-$ in water droplets simulated at temperature of 300 K have been taken from ~\citeauthor{bouchafra2018predictive}\cite{bouchafra2018predictive}, and originate from classical molecular dynamics (CMD) simulations employing polarizable force fields \cite{real2016structural}. Here we have considered the same 100 snapshots as in the original reference. Each droplet contains 50 water molecules, and the halogen position has been constrained to be at the center of mass of the system.

Initial structures of the halogens adsorbed on the ice surfaces have been taken from~\citeauthor{woittequand2007classical}\cite{woittequand2007classical}, which are based on CMD simulation of HCl adsorbed on the ice surfaces with a non-polarizable force field~\cite{toubin2002structure} at 210 K. We have considered 25 snapshots, each containing 216 water molecules. It should be noted that this set of structures account for the disorder at the air-ice interface associated with a thin ice quasi-liquid layer (QLL). 

Due to the nature of the force field, structures for Cl$^-$ were not available for the same surface, and we have therefore started out from the HCl snapshots, removed a proton and proceeded to optimizations of the ion position while constraining the water molecules of the ice surface to keep their original positions. As such, the adsorption site is sightly altered with respect to the original HCl-ice system, but not the surface on which adsorption takes place.

Furthermore, to assess the importance of HCl-water interactions not captured by the classical force field, we have applied a constrained optimization to the HCl species as well, in a similar vein as outlined above, for all CMD snapshots. We have considered two situations: one in which only the position of HCl was allowed to change (thus allowing both changes in H-Cl bond distance and in relative position of H and Cl with respect to the surface), and another in which the atoms for the six waters closest to HCl were also allowed to change position.

We note that considering the charged system without a counter ion implicitly assumes a model for a diluted solution/interface. From prior work with charged species in the literature such an approach is warranted if one takes into account the effects of the polarization of the solvent~\cite{Yu2010,Videla2015,real2016structural,Houriez2019,Acher2020,Li2021,Berkowitz2021}, and has been shown to yield spectroscopic results~\cite{Skanthakumar2017,bouchafra2018predictive} that closely match experimental measurements, provided of course the model mirrors the essential features of the experimental system.

\subsection{Embedding models}

\begin{figure*}[htp!]
\centering
\includegraphics[width = 18.0cm, trim = 0cm 0cm 0cm 0cm]{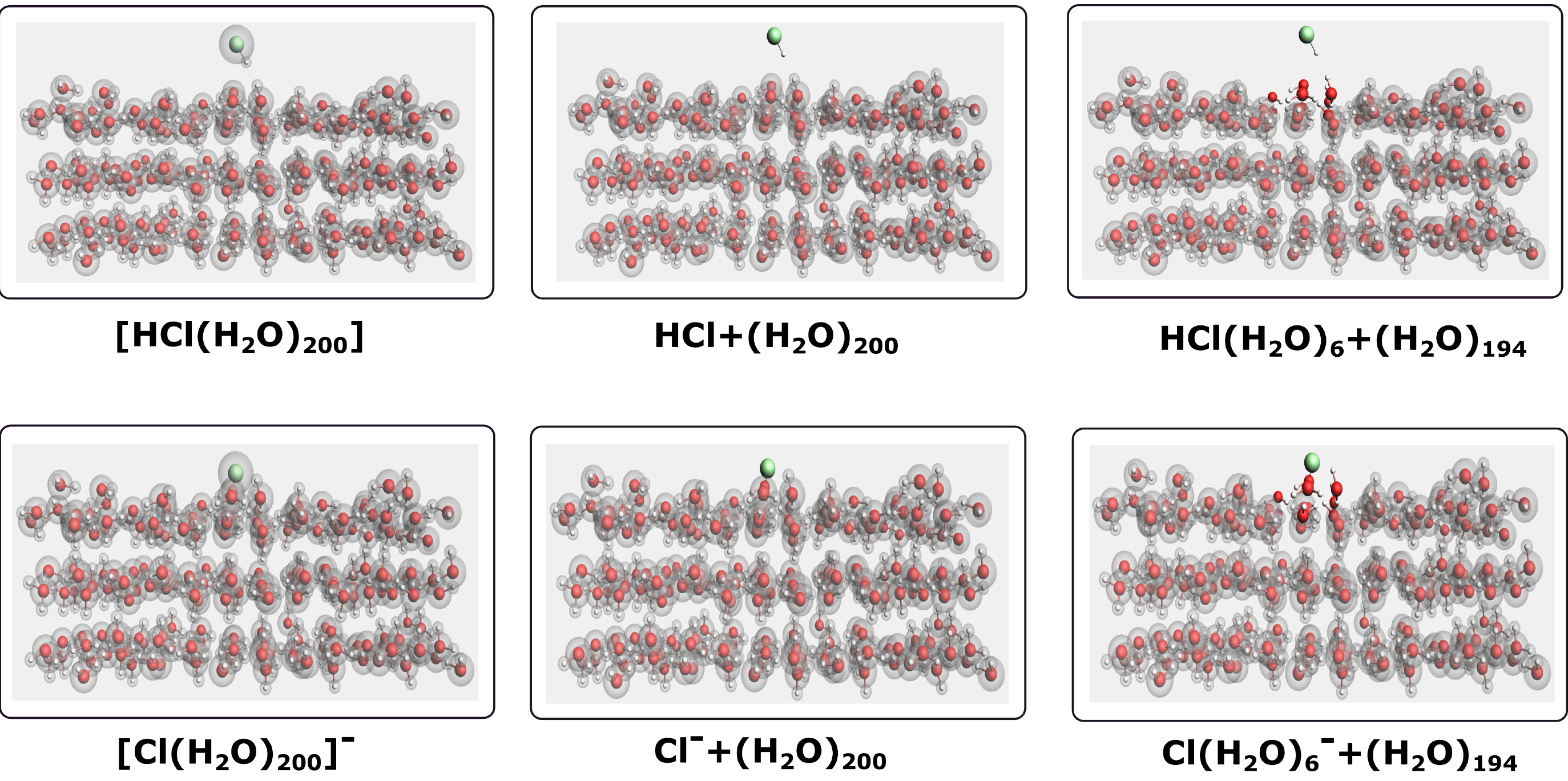}
\caption{\label{fig:modelstofollow}Perspective views for cluster models of halogens adsorbed ice surfaces (represented by 200 water molecules): without a partition into subsystem (\textbf{SM}, left)), with an active subsystem containing only the halogen (\textbf{EM1}, center) and with an active subsystem containing the halogen and 6 nearest water moleculues (\textbf{EM2} right). Boxes A, B and C represent the system HCl-ice whereas boxes D, E and F represent the system Cl$^-$-ice.}
\end{figure*}

For Cl$^-$ in water droplets a single embedding model is used, in which two subsystems are defined: the active subsystem (treated with CC), containing the halogen, and the environment (treated with DFT) composed of the 50 waters. Further details can be found in~\citeauthor{bouchafra2018predictive}~\cite{bouchafra2018predictive}. For our discussion of the halogens adsorbed on ice we have considered three models: the first represents calculations without embedding (referred to as \textbf{SM}, for supermolecular model, in what follows). Furthermore, two embedding models are considered, the first (referred to as model \textbf{EM1} in the following) is similar to the droplet one in that  only the species containing the halogen is contained in the active subsystem, and all water molecules make up the environment. In the second model (referred to as model \textbf{EM2} in the following), we include a number of water molecules (the nearest neighbors to the halogen species) in the active subsystem, and the remaining water molecules make up the environment. These models are pictorially represented in figure~\ref{fig:modelstofollow}. In the following, \textbf{EM2$^{V}$} will denode a model containing one water molecule in the active subsystem. Additional details can be found in the supplementary information. 

\section{Results and Discussion\label{sec:discussion}}

Before proceeding to the discussion of our results for embedding systems, we consider it instructive to address first the performance of theoretical approaches for obtaining core binding energies for the isolated (gas-phase) species, since these provide us with a well-defined reference point with which to assess the behavior of embedded models in complement to a comparison to experimental results~\cite{parent2011hcl, kong2017coexistence}.

\subsection{\label{sec:gas-phase-results} Gas-phase calculations}

Our results for gas-phase calculations are shown in table~\ref{tab:ham_bas}. Considering first the EOM calculations, our choice of focusing on the $^2$DCG$^M$ Hamiltonian stems from the fact that the Gaunt interaction is essential for properly describing the K edge, while showing non-negligible effects for the L edges (see supplementary result for a comparison to two- and four-component results based on the Dirac-Coulomb Hamiltonian). Here, we also provide an investigation of basis set convergence with extended basis sets (including an extrapolation to the complete basis set limit), as well as a in-depth comparison to experimental results. 
For the SAOP model potential, on the other hand, we are not aware of any comparison for core ionization energies employing the equivalent of Koopmans theorem for DFT~\cite{Chong2002}--that is, by obtaining the core binding energy as the negative of the Kohn-Sham orbital energies (BE = $-\epsilon_\text{KS}$). A Koopmans' approach will be inherently less accurate for core electron binding energies due to the lack of relaxation effects~\cite{PueyoBellafont2015}, though it remains very convenient for a qualitative understanding of chemical shifts for a particular edge. In the case of SAOP, which has shown a very good performance for valence ionizations of halides and water in droplets~\cite{bouchafra2018predictive}, it is interesting to verify by how much it deviates from EOM calculations for deeper ionizations.

The presentend SAOP binding energies correspond to those obtained with the scaled (spin-orbit) ZORA Hamiltonian (we recall that while scaled and unscaled ZORA energies differ significanly for deep cores, the eigenfunctions for both cases are the same~\cite{vanLenthe1994}). A comparison of SO-ZORA SAOP and $^2$DCG$^M$ EOM results shows, unexpectedly, a marked difference between the K, L$_1$, and L$_2$/L$_3$ edges. For L$_2$/L$_3$ edges of both HCl and Cl$^-$ the difference between methods is reughly 16 eV, increases to roughly 27 eV for the  L$_1$, and reaches around 70 eV for the K edge, which represent differences in binding energies of roughly 8-9\%, 10\% and 2-3\% for the respective edges. That the difference between SO-ZORA results and EOM is essentially the same for the two species will be useful for the comparison between structural models that follows. 

From the table, we observe that our triple-zeta basis $^2$DCG$^M$ EOM results overestimate the experimental  L$_2$ and L$_3$  gas-phase HCl binding energies reported by~\citeauthor{hayes1972absorption}~\cite{hayes1972absorption} and~\citeauthor{aitken1980electron}~\cite{aitken1980electron} by around 1 eV. A similar difference is observed for the L$_1$ edge of chloride in the NaCl crystal, with the L$_2$ and L$_3$ edges in this case showing deviations smaller than 1 eV. For the K edge there appears to be a slightly larger discrepancy (around 1.7 eV) between theory and the experimental values quoted by~\citeauthor{thompson2009xps}~\cite{thompson2009xps}. In spite of the fact that for all edges these experimental results are not for a gas-phase chloride atom, we take the overall very good agreement to indicate that $^2$DCG$^M$ EOM should show an uniform accuracy for both species. 
 
\begin{table*}[htp]
\centering
\caption{\label{tab:ham_bas} CVS-EOM-IP-CCSD chlorine core binding energies (in eV) for HCl and Cl$^{-}$ in gas-phase for the $^2$DCG$^M$ Hamiltonian and employing triple-zeta basis sets as well as values extrapolated to the complete basis set limit (CBS). In addition to those, we present core binding energies obtained via the analogue of Koopmans theorem for DFT~\cite{Chong2002}, employing the SAOP model potential for the ZORA Hamiltonian. In parenthesis we presente the differences with respect to the CVS-EOM-IP-CCSD $^2$DCG$^M$ results with triple-zeta basis sets, which we take as reference. Apart from the energies for the individual edges, we provide the core binding energy shift ($\Delta_{BE}$, in eV) between HCl and Cl$^{-}$ for the theoretical gas-phase values.} 
\begin{tabular}{l l l r r r r}
\hline
\hline
Species & Hamiltonian & Method & K & L$_1$ & L$_2$  &L$_3$ \\
\hline
HCl      & SO-ZORA     & SAOP   & 2764.58 (-69.32) & 253.94 (-26.75) & 194.08 (-15.94) & 192.41 (-15.98) \\
         & $^2$DCG$^M$ & EOM    & 2833.90 (  0.00) & 280.69 (  0.00) & 210.02 (  0.00) & 208.39 (  0.00) \\
         & $^2$DCG$^M$, CBS& EOM& 2833.86 ( -0.04) & 280.79 (  0.10) & 210.18 (  0.16) & 208.55 (  0.16) \\
         & Exp. (gas phase)~\cite{hayes1972absorption} &&&& 208.70 & 207.1 \\
         & Exp. (gas phase)~\cite{aitken1980electron}  &&&& 209.01 & 207.38 \\
         & & & & & \\
Cl$^{-}$ & SO-ZORA     & SAOP   & 2754.42 (-71.63) & 243.88 (-26.93) & 184.02 (-16.24) & 182.35 (-16.21) \\
         & $^2$DCG$^M$ & EOM    & 2824.17 (  0.00) & 270.73 (  0.00) & 200.10 (  0.00) & 198.47 (  0.00) \\
         & $^2$DCG$^M$, CBS& EOM& 2824.13 ( -0.04) & 270.84 (  0.11) & 200.29 (  0.20) & 198.66 (  0.19) \\
         & Exp. (NaCl)~\cite{crist2019xps} &   &  & 269.6          & 200.6          & 198.9           \\
         & Exp. (KCl)~\cite{moulder1995xps} &  &  &                & 200.1          & 198.6           \\
         & Exp. (NaCl)~\cite{thompson2009xps} && 2822.4 & 270  & 202 & 200  \\
         & & & & & \\
$\Delta_{BE}$& SO-ZORA     & SAOP   & 10.16 & 10.06 & 10.06 & 10.06 \\ 
             & $^2$DCG$^M$ & EOM    & 9.73  & 9.96 &  9.92 & 9.92 \\
         & $^2$DCG$^M$, CBS& EOM    & 9.73  & 9.95 & 9.89 & 9.89 \\ 
\hline
\hline
\end{tabular}
\end{table*}

We note that there remain three potential sources of errors in our calculations: (a) basis set incompleteness; (b) QED and retardation effects; and (c) energy corrections due to higher-order excitations in the CC wavefunctions. For HCl there could be a fourth, the H-Cl bond distance, but as can be seen from results of a scan of this coordinate (see supplementary information), even large variation of bond lengths don't change energies by more than a 0.1-0.2 eV for the L edges, and around 0.3 eV for the K edge, meaning that the first three factors should be behind most of the discrepancy.

From the literature~\cite{Yerokhin2017}, QED and retardation effects are expected to be well below 0.1 eV for chlorine.  We are unable at this point to deternine the effect of higher-order excitations with the current implementation in DIRAC.  For assessing the basis set effects, we have calculated binding energies with quadruple zeta basis sets and with them and the triple-zeta results obtained  the complete basis set limit (CBS) values shown in table~\ref{tab:ham_bas}.  

Comparing the EOM CBS results to the experimental results of~\citeauthor{hayes1972absorption}~\cite{hayes1972absorption} and~\citeauthor{aitken1980electron}~\cite{aitken1980electron} for the L$_2$ and L$_3$ edges of HCl, we see that agreement gets slighty worse than with the triple-zeta basis, and an interesting point is that for the K edge the effect of extrapolation is to reduce the binding energies whereas for the L edges the opposite is true. For Cl$^-$, we see essentially the same variation between the triple zeta and the extrapolated values as for HCl. In both cases, as the L$_2$ and L$_3$ edges are affected by the same amounts, the spin-orbit splitting of the 2p$_{1/2}$ and 2p$_{3/2}$ remains at around  1.6 eV for both species, a value consistent with the gas-phase experimental values for HCl (in the NaCl crystal, this splitting is of 1.7 eV~\cite{crist2019xps} whereas in KCl it is 1.6 eV~\cite{moulder1995xps}). 

Finally, we see that the $\Delta_{BE}$ value are nearly indistinguishable, in spite of the calculations having employed different Hamiltonians or correlated methods. This indicates that this is a robust quantity to (semi-quantitatively) characterize the chemical shift in the binding energies, even when the absolute binding energies are rather poor (as is the case of DFT obital energies). 

\subsection{\label{sec:comparison-supermolecule}Assessing the embedded models}

Before proceeding to the calculation of halide binding energies considering the sampling of configurations from MD, it is necessary to make an initial assessment of the quality of the embedding methods, as well as the suitability of the structural models to which we apply the embeding methods. In the first case, the most straightforward evaluation comes from comparing how our embedding models (EM1 and EM2) can reproduce the reference supermolecular model (SM); in the second case, we shall be interested on characterizing the long-range interaction between the halogenated species and its environment. 

As discussed in sec.~\ref{sec:gas-phase-results}, the ZORA/SAOP model is sufficiently systematic to allow us to compare the behavior of both embedded HCl and Cl$^-$ by using orbital energies as proxies for the extent to which our embedding models (EM1 and EM2) can reproduce the reference supermolecular model (SM) for the different core edges. From  prior work~\cite{gomes2008calculation,actinide-Gomes-PCCP2013-15-15153,bouchafra2018predictive} we expect the need for subsystem DFT calculations (in which both subsystem densities are optimized via freeze-thaw cycles) for  Cl$^-$, whereas for the neutral HCl FDE calculations (in which the electron density for the environment--the ice surface--is constructed in the absence of the halogen species)  relaxation of the environment would be less of an issue~\cite{gomes2008calculation,rt_fde2020}, and therefore little should be gained from subsystem DFT calculations. 

\begin{figure*}
\centering
\includegraphics[width=\linewidth]{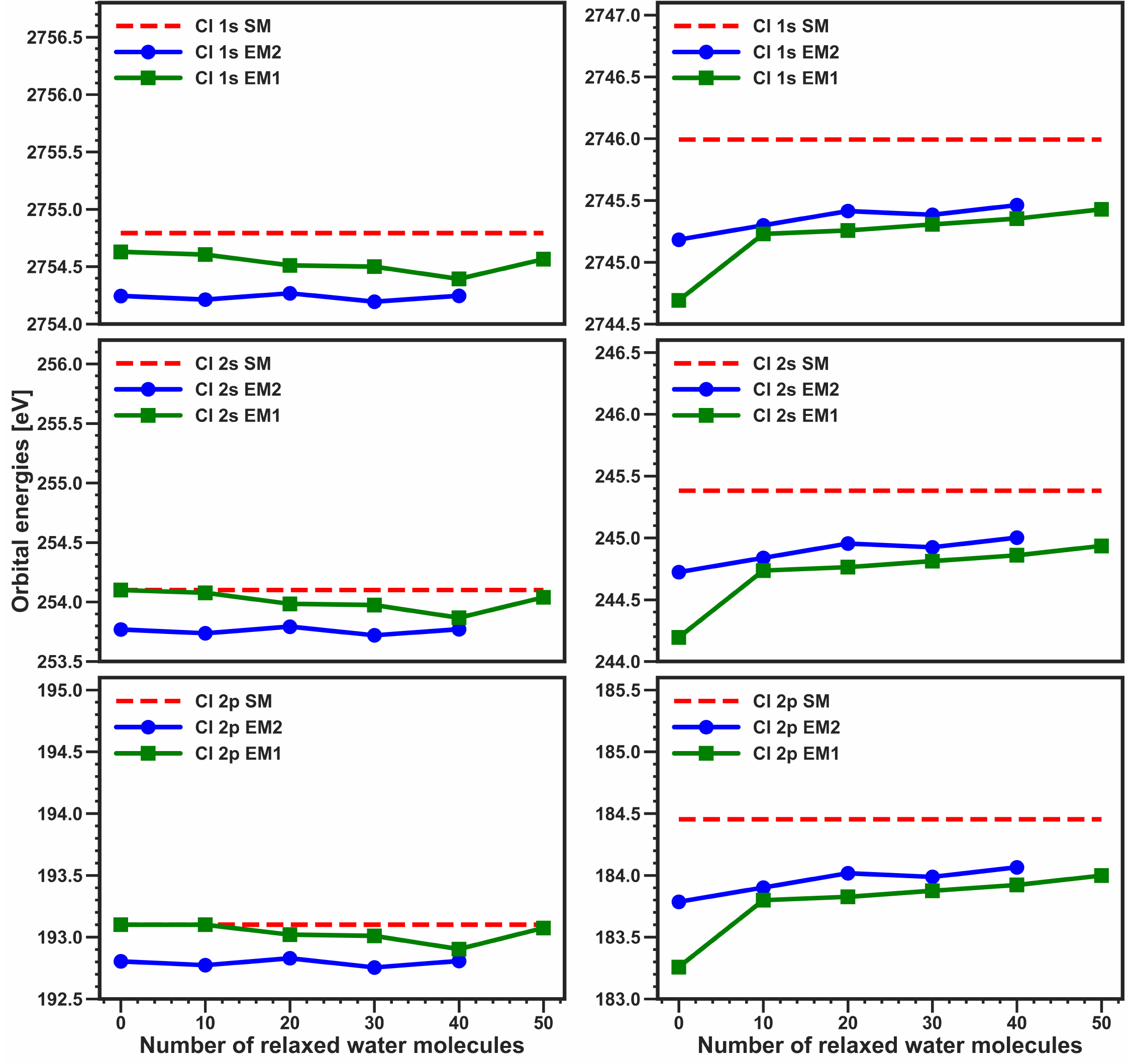}
\caption{\label{fig:water-binding-energy-freeze-thaw} Variation of the approximate K, L$_1$ and L$_{2,3}$ core binding energies of chlorine in HCl (left) and Cl$^-$ (right), obtained from scalar ZORA calculations with the SAOP model potential, with respect to the number of water molecules whose density is relaxed (in the ground state) via freeze-thaw cycles, for models \textbf{EM1} (squares) and \textbf{EM2} (circles). For comparison, the corresponding orbital energies obtained for model \textbf{SM} are provided as a reference (dashed line). The L$_{2,3}$ values are not split as calculations do not include spin-orbit coupling.}
\end{figure*}

In figure~\ref{fig:water-binding-energy-freeze-thaw} we present the comparison, for a selected snapshot, of the different models and how the core binding energies vary as the number of water molecules is included in the Freeze-thaw procedure (for EM1 and EM2 we relax at most the 50 and 40 water molecules nearest to the halogen, respectively).  The first important difference between the HCl and Cl$^-$ systems is that, as expected, for  Cl$^-$ there is a much more important change between FDE calculations (zero relaxed water molecules) and subsystem DFT (at least one relaxed water molecule) ones, with FDE calculations with the EM1 model showing discrepancies of around 1.3 eV from SM for all edges.

By adding the six nearest water molecule to the active subsystem in EM2, we observe the difference to SM for FDE is reduced to around 0.8 eV. When 10 water molecules are relaxed, the subsystem DFT calculations with EM1 yield roughly the same results as FDE ones for EM2, and from this point onwards both EM1 and EM2 subsystem DFT calculations yield binding energies that differ by around 0.1 eV at most. It is important to note that even after 50 (40) water molecules relaxed, EM1 and EM2 still underestimate the SM binding energies by about 0.5 eV for the L edges and 0.6 eV for the K edge. 

For HCl, on the other hand, we see very little improvement over FDE for the subsystem DFT calculations: varying the number of relaxed water molecules from zero to 40 or 50 introduces variations of at most 0.1-0.2 eV. This is in line with our expectation that FDE already provides a rather good representation of the environment for neutral subsystems. In contrast to Cl$^-$, we observe that EM2 shows slightly worse agreement to SM than EM1, and that for the L edges the embedding approaches are about 0.1-0.2 eV closer to the reference SM binding energies than for the K edge. In spite of that, embedded models still show a rather good agreement to SM, with EM2 typically underestimating the SM binding energies by 0.3 eV for the L edges (and 0.5 eV for the K edge), whereas EM1 reproduces SM binding energies nearly exactly for the L edge while underestimating the K edge binding energies by around 0.2 eV

Taken together, these results make us confident that, first and foremost, embedded models (and consequently, the underlying embedding potential) are indeed capable of reproducing SM calculations, and to do so in a manner that is roughly uniform for the K and L edges alike. 

Second, that subsystem DFT derived embedding potentials introduce errors (due to the limited accuracy of the approximate kinetic energy density functionals employed to calculate the non-additive kinetic energy contributions~\cite{jacob2014subsystem}) that should result in small but non-negligible (0.5 eV or lower) underestimation of SM DFT binding energies. 

In view of using them in CC-in-DFT calculations, the EM2 model has a significant disadvantage in that the active subsystem is significanly larger than EM1, and given the small difference in performance between the two, here we have opted to focus from now on on the EM1 model, and employing the EM2$^v$ model (that contains only one water molecule in the active subsystem instead of six) whenever assessing the suitability of the EM1 model in CC-in-DFT calculations. 

\begin{figure*}[htp]
\includegraphics[width=\linewidth]{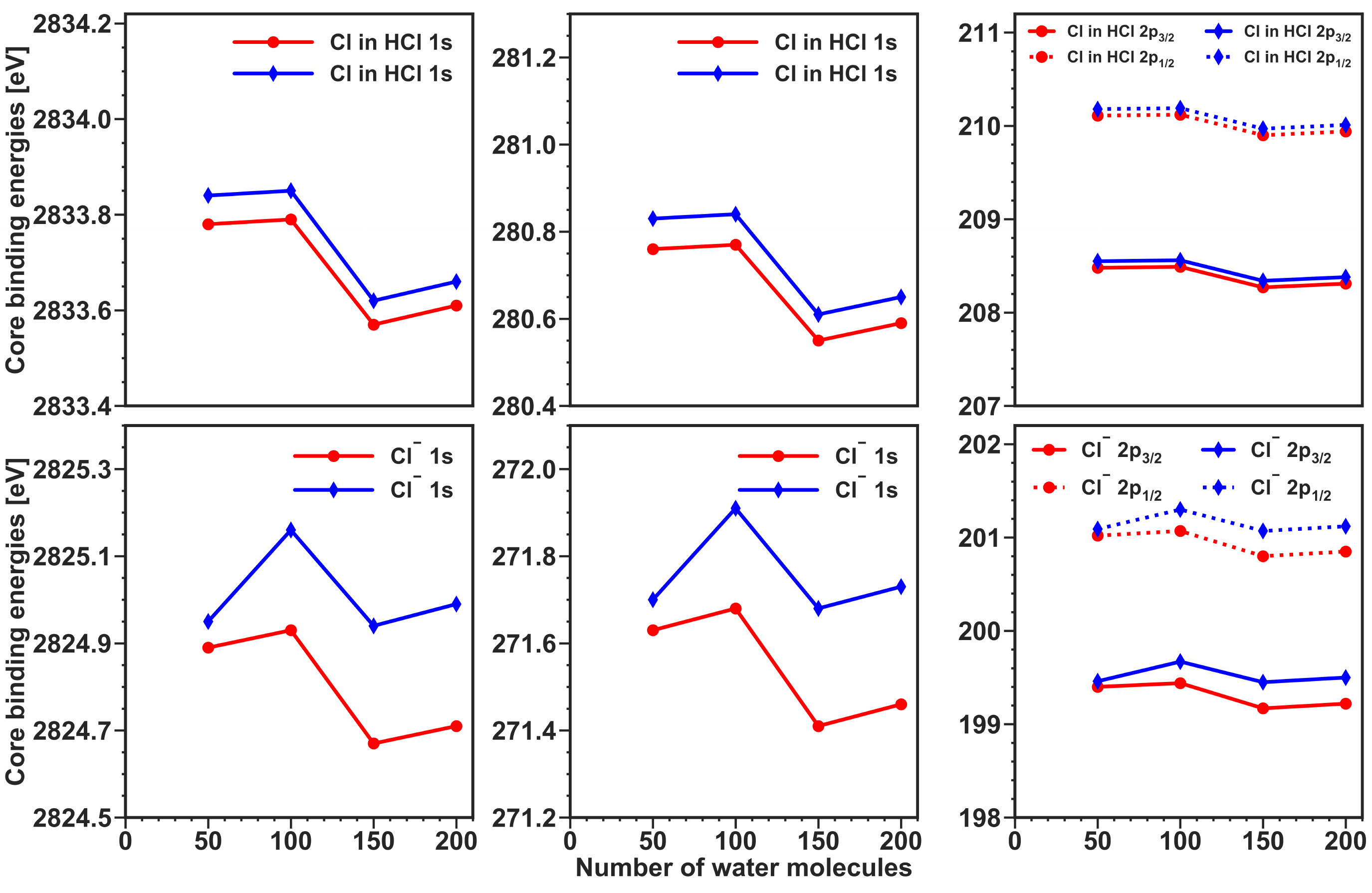}
\caption{\label{fig:eom-cl-hcl-core} CC-in-DFT K, L$_1$, L$_2$ and L$_3$ triple-zeta binding energies of  HCl and Cl$^{-}$ adsorbed on ice surfaces for a single snapshot, as a function of the number of water molecules in the environment (in addition to the 50 molecules nearest to the halogen system that are always taken into account). Blue lines represent the system with only the halogen species in the active subsystem, and red lines active systems containing halogen species and one explicit water molecule.}  
\end{figure*}

In figure~\ref{fig:eom-cl-hcl-core} we employ the EM1 and EM2$^v$ models, again for a single snapshot (and therefore disregard temperature effects introduced by considering several snapshots, as it will be done in the following), to verify the effect of long-range interactions between the halogens and the ice, through the truncation of the size of the water environment in the CC-in-DFT calculations.

We observe that for HCl there is no discernible difference between the embedded models, and that long-range effects seem to represent relatively small (0.2 eV) contributions, that are roughly uniform for the different edges, and tend to lower the core binding energies. Interestingly, the plots seem to indicate that long-range effects start to kick in after more than 100 water molecules have been taken into account. For Cl$^-$, the situation is qualitatively slightly different, since we see a non-negligible difference between the EM1 and EM2$^v$ CC-in-DFT results, with the latter showing binding energies typically 0.2 eV lower than the former. That said, there appears to be a small decrease in binding energies between 100 and 150 water molecules (0.2 eV), as seen for HCl. We also note that irrespective of the model (EM1 and EM2$^v$), the splitting between L$_2$ and L$_3$ edges remains around 1.6 eV.

Further evidence of the relative insensitivity of the results to the size of the cluster beyond 50 water molecules is given by the analysis of the system's dipole moment as a function of number of water molecules, as shown in table S4 in the supplementary materials, in that there are no significant changes in dipole for sufficiently large system.  

From these results, we consider that employing the EM1 model is still advantageous from a computational point of view, since we consider that its smaller computational cost offsets the relatively modest improvement brought about by explicitly considering a water molecule in the active subsystem. Furthermore, due to the small changes upon considering a much larger environment, for the following we shall only consider models containing the halogen system and 50 water molecules.

\subsection{\label{sec:configurational-averaging}Configurational averaging : ice and droplet models}

Having established above that the EM1-based model containing 50 water molecules provides a very good balance between the ability to faithfully reproduce the reference calculations and the computational cost associated with CC-in-DFT calculations, we now turn to a discussion of the effects of the structural model for the environment and of the temperature, both associated with considering snapshots from classical molecular dynamics simulations. Table ~\ref{tab:ccsd-ice-hcl} summarizes our results.

Starting with the chloride ion in a droplet we observe that our calculated triple-zeta quality binding energies (BE(A), calculated from the average of binding energies over 100 snapshots of a simulation at 300K) show a slighly larger shift with respect to the gas-phase value ($\Delta^g$BE(A)) for the K edge (around 6.7 eV) than for the L edges (around 5.9 eV), which is consistent with the picture from our analysis of the single snapshot ZORA/SAOP results in section~\ref{sec:comparison-supermolecule}. We have not carried out quadruple zeta calculations for this system, due to the fact that the CBS corrections to the triple-zeta values for the chloride-ice surface (see supplementary information), at least for the L edges, are rather similar to the ones obtained for the gas-phase system. As such, we have decided to apply the gas-phase corrections for both K and L edges to the droplet system, given that for valence ionizations~\cite{bouchafra2018predictive} CBS corrections from gas-phase or from averaged droplet binding energies were essentially the same.

A comparison of the droplet CBS-corrected values to the experimental results of~\citeauthor{Pelimanni2021}~\cite{Pelimanni2021}, which have measured the L$_2$ and L$_3$ edges for KCl solutions at different concentrations and at somewhat higher temperatures (nozzle temperature of 373K), shows our results are in good agreement with experiment, as our results overestimate experiment by almost exactly 1 eV for each edge. 

Our 2p spin-orbit splitting is consistent with that of experiment, at around 1.6 eV, a value that is close to the one seen in the gas phase (roughly the same differences are found in comparison to the experimental results of~\citeauthor{Partanen2013}~\cite{Partanen2013}). There are much more significant discrepancies between our simulations and the experimental results of~\citeauthor{kong2017coexistence}~\cite{kong2017coexistence} obtained at somewhat lower temperatures (253K), not only in terms of the binding energy values (which are around 4 eV lower than our results) but of the 2p spin-orbit splitting (2.1 eV), which is 0.5 eV larger than both our simulation and other experimental results~\cite{Partanen2013,Pelimanni2021}.

The discrepancy between our results and those of~\citeauthor{kong2017coexistence} could be due to temperature effects, since encapsulation of the halogens is driven by the temperature induced surface disorder, though the role of other parameters such as differences in calibration in the BE scale, cannot be dismissed out of hand. We note that \citeauthor{kong2017coexistence} have used as internal reference the oxygen K edge, but since obtaining such data is beyond the scope of this work (as it would require the construction and validation of new embedding models in order to carry out CC-in-DFT calculations on water molecules), we provide in Table ~\ref{tab:ccsd-ice-hcl} results for the valence band of water for the droplet model, obtained from SAOP orbital energies by~\citeauthor{bouchafra2018predictive}~\cite{bouchafra2018predictive}. 

As discussed in the original work, the SAOP valence band of water is quite consistent with the available experimental results, with the 1 eV underestimation of experimental values having to do with the finite size of the water droplet~\citeauthor{bouchafra2018predictive}. The good agreement between our theoretical values and the experimental values of~\citeauthor{Pelimanni2021} for the valence band of water, but also for the valence band of chloride, make us confident in the reliability of our embedded models and the CC-in-DFT protocol for core edges.

 \begin{table*}
\centering
\caption{\label{tab:ccsd-ice-hcl} Mean values of CC-in-DFT chlorine core binding energies (BE, in eV) averaged over structures from CMD simulations for models with 50 water molecules, and the difference of BEs and those calculated for the gas phase ($\Delta^g$BE, in eV). The molecular structures correspond to (A) the original CMD snapshots for water droplets~\cite{bouchafra2018predictive} and ice surfaces~\cite{woittequand2007classical}; (B) optimization of the halogen system coordinates, keeping the ice surface constrained to the CMD structure; (C) optimization of the halogen system coordinates and four nearest neighbor waters, keeping the remaining of the ice surface constrained to the CMD structure. For the ice surface systems, calculations correspond a temperature of 210 K. For water droplets, calculations correspond a temperature of 300 K. We also provide theoretical results (scalar ZORA SAOP) for valence bands (3a$_1$ and 1b$_1$) of water for the ice surface, and for completeness we also provide for the water droplets the CC-in-DFT chlorine 3p and SAOP water 3a$_1$ and 1b$_1$ BEs from~\citeauthor{bouchafra2018predictive}~\cite{bouchafra2018predictive}, obtained for the same snapshots as the chlorine BEs. We compare these results to experimental results by~\citeauthor{kong2017coexistence}~\cite{kong2017coexistence} (253 K),~\citeauthor{parent2011hcl}~\cite{parent2011hcl} (90 K),~\citeauthor{Partanen2013}~\cite{Partanen2013} (393-423 K),~\citeauthor{Pelimanni2021}~\cite{Pelimanni2021} (373 K),~\citeauthor{Kurahashi2014}~\cite{Kurahashi2014} (280 K) and~\citeauthor{Winter2005}~\cite{Winter2005}.}
\begin{tabular}{l rrrrrrrr p{0.1cm} r r}
\hline
\hline
&&\multicolumn{6}{c}{triple-zeta results} & & & \\
\cline{4-9}
System   & Environment
         & ionisation
         & BE(A) & $\Delta^g$BE(A)
         & BE(B) & $\Delta^g$BE(B)
         & BE(C) & $\Delta^g$BE(C)
         && BE(CBS$^\dagger$)
         &  Experiment \\
\hline
 HCl     & ice     & K      & 2834.03 & 0.13 & 2833.40 &-0.50 & 2833.29 &-0.61 &&  2834.33 & 2817.6~\cite{kong2017coexistence} \\
         &         & L$_1$  &  280.84 & 0.15 &  280.15 &-0.54 &  280.04 &-0.65 &&   281.19 &       \\
         &         & L$_2$  &  210.18 & 0.15 &  209.49 &-0.53 &  209.38 &-0.64 &&   210.60 &  204.9~\cite{kong2017coexistence} \\
         &         &        &         &      &         &      &         &      &&          &  202.2~\cite{parent2011hcl} \\
         &         & L$_3$  &  208.54 & 0.15 &  207.86 &-0.53 &  207.75 &-0.64 &&   209.04 &  202.8~\cite{kong2017coexistence} \\
         &         &        &         &      &         &      &         &      &&          &  200.9~\cite{parent2011hcl} \\
&&&&&&&&&&& \\
Cl$^{-}$ &         & K      &         &      & 2827.70 & 3.53 &         &      &&  2828.79 &  2815.4~\cite{kong2017coexistence}\\
         &         & L$_1$  &         &      &  274.90 & 4.17 &         &      &&   275.07 &         \\
         &         & L$_2$  &         &      &  204.41 & 4.31 &         &      &&   204.47 &   202.7~\cite{kong2017coexistence}\\
         &         &        &         &      &         &      &         &      &&          &   199.6~\cite{parent2011hcl} \\
         &         & L$_3$  &         &      &  202.70 & 4.23 &         &      &&   202.91 &   200.6~\cite{kong2017coexistence}\\
         &         &        &         &      &         &      &         &      &&          &   198.3~\cite{parent2011hcl} \\
         &         & 3p     &         &      &         &      &         &      &&          &     10~\cite{parent2011hcl} \\
&&&&&&&&&&& \\
Cl$^{-}$ & droplet & K      & 2829.97 & 6.66 &         &      &         &     &&   2829.93 &   \\
         &         & L$_1$  &  276.63 & 5.90 &         &      &         &     &&    276.74 &   \\
         &         & L$_2$  &  205.99 & 5.89 &         &      &         &     &&    206.19 &    205.0~\cite{Pelimanni2021}\\
         &         &        &         &      &         &      &         &     &&           &    205.0~\cite{Partanen2013}\\ 
         &         &        &         &      &         &      &         &     &&           &    202.7~\cite{kong2017coexistence}\\ 
         &         & L$_3$  &  204.36 & 5.89 &         &      &         &     &&    204.55 &    203.4~\cite{Pelimanni2021}\\
         &         &        &         &      &         &      &         &     &&           &    203.4~\cite{Partanen2013} \\
         &         &        &         &      &         &      &         &     &&           &    200.6~\cite{kong2017coexistence}\\ 
         &     & 3p$_{1/2}$ &    9.9~\cite{bouchafra2018predictive} & & & & & &&     10.1~\cite{bouchafra2018predictive}  &       \\
         &     & 3p$_{3/2}$ &    9.7~\cite{bouchafra2018predictive} & & & & & &&      9.9~\cite{bouchafra2018predictive}  
                                                                                          & 9.8~\cite{Pelimanni2021} \\
         &         &        &         &      &         &      &         &     &&          & 9.5~\cite{Kurahashi2014} \\
         &         &        &         &      &         &      &         &     &&          & 9.6~\cite{Winter2005} \\
&&&&&&&&&&& \\
H$_2$O   & Cl$^-$ ice&3a$_1$& 16.1    &      &         &      &         &     &&          & 13.7~\cite{parent2011hcl} \\
         &         & 1b$_1$ & 12.6    &      &         &      &         &     &&          & 12~\cite{parent2011hcl}  \\
&&&&&&&&&&& \\
         & droplet & 3a$_1$ &   12.5~\cite{bouchafra2018predictive}  &&&&&    &&          &   13.76~\cite{Partanen2013} \\
         &         &        &         &      &         &      &         &     &&          &   13.78~\cite{Kurahashi2014} \\
         &         &        &         &      &         &      &         &     &&          &   13.50~\cite{Winter2005} \\ 
         &         & 1b$_1$ &   10.4~\cite{bouchafra2018predictive}  &&&&&    &&          &   11.4~\cite{Pelimanni2021} \\
         &         &        &         &      &         &      &         &     &&          &   11.41~\cite{Partanen2013} \\
         &         &        &         &      &         &      &         &     &&          &   11.31~\cite{Kurahashi2014} \\
         &         &        &         &      &         &      &         &     &&          &   11.16~\cite{Winter2005} \\
\hline
\hline
\end{tabular}
\end{table*}

Considering now the chloride ion at the ice surface, our  calculated triple-zeta quality binding energies (BE(B), calculated from the average of binding energies over 25 snapshots obtained by a combination of MD simulations at 210K for HCl, followed by an optimisation of the chlorine atom position) show similar behavior to that of droplets with respect to the free ion, with $\Delta^g$BE(B) values which are nearly the same for the K and L edges, again in line with the picture that the embedding potentials for these calculations affect the K and L edges in a roughly homogeneous manner. Here, however, we have a smaller shift for the K edge (3.53 eV) than for the L edges, and for the latter the differences between L$_1$, L$_2$ and L$_3$ are of the order of 0.1 eV, that is, an order of magnitude more than for the water droplet. 

In qualitative terms, our calculations reproduce well the trend of decreasing binding energies when going from solution--represented by the experimental results of~\citeauthor{Pelimanni2021}~\cite{Pelimanni2021}--to the surface--represented by the experimental results of~\citeauthor{parent2011hcl}~\cite{parent2011hcl} for lower temperatures and ~\citeauthor{kong2017coexistence}~\cite{kong2017coexistence} for higher temperatures (though for the latter, a near equivalence between the reported chloride binding energies for NaCl solution and ice surface 253K would indicate that chloride does behave as a free ion in both systems).

Quantitatively, our CBS corrected triple zeta results show differences of the order of 4 eV for the L$_2$ and L$_3$ edges with respect to the experimental results of~\citeauthor{parent2011hcl}~\cite{parent2011hcl}, obtained at 90K. At the same time the water valence band for the theoretical model for the ice surface is in good agreement with the same low-temperature experiment, with discrepancies of around 0.6 eV for the 1b$_1$ band. For the 3a$_1$ band discrepacies are of about 2 eV, but experimentally that is a broader band and therefore more difficult to provide an unambiguous comparison between theory and experiment. The difference of performance of our models for the ice surface valence binding energies and the core chloride binding energies could be an indication of the importance of temperature effects, that cannot be properly accounted for in our models since we only have data for 210K. 

The discrepancies for core BEs are smaller with respect to the results of~\citeauthor{kong2017coexistence}~\cite{kong2017coexistence}, measured at 253K (and thus closer to the simulation conditions), but our values still overestimate the experimental results by 1.77 eV and 2.31 eV for the L$_2$ and L$_3$ edges, respectively. This is larger than the differences we observe for droplets between theory and the results of~\citeauthor{Pelimanni2021}~\cite{Pelimanni2021}, but somewhat smaller than the differences between our droplet model and the results for NaCl solution from~\citeauthor{kong2017coexistence}~\cite{kong2017coexistence}. 
In our view, taken together the results for Cl$^-$ on ice and water droplets seem to indicate a fairly systematic difference between our theoretical models and the experimental results of~\citeauthor{kong2017coexistence}. 

\begin{figure*}[htp!]
\centering
\includegraphics[width=\linewidth]{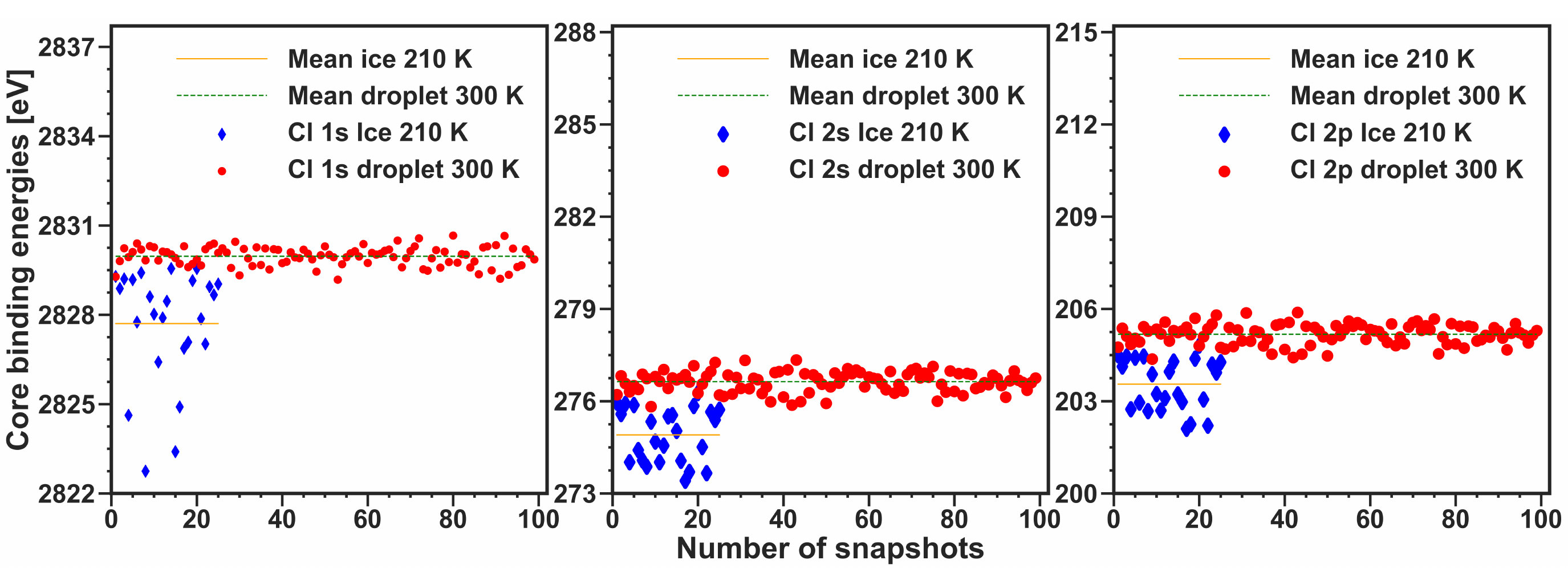}
\caption{\label{fig:eom-cl-hcl-average300-210} CC-in-DFT K, L$_1$, L$_2$ and L$_3$ triple-zeta binding energies of chloride adsorbed on ice surfaces at 210 K (results averaged over 25 snapshots) and in water droplets at 300 K (results averaged over 100 snapshots).}
\end{figure*}

On the simulation side, there is an important difference between the chloride-ice system with respect to the droplets, which is linked to the process and quality of the sampled structures, since sampling is intimately connected to the description of temperature effects. By inspecting figure~\ref{fig:eom-cl-hcl-average300-210}, in which we show the K and L edge binding energies obtained for each snapshot  around the mean value presented on table~\ref{tab:ccsd-ice-hcl}, we see narrow distributions around the mean for the droplet system for all edges considered (within envelopes of around 1 to 1.5 eV). For the chloride-ice system, the distributions are much wider, and of around 3 eV for the L edges, and almost 10 eV for the K edge.
 
This difference is in part expected, since in the droplet model the chloride ion is always completely surrounded by water molecules--and therefore one can consider that on average the ion has always a fairly constant degree of interaction with its environment--whereas for the ice model, the amount of water molecules with which the ion interacts greatly depends upon how much it has penetrated into the QLL. We speculate that, in our case, the sampled structures place the chloride ion deeper than it would be on average, and with that our results could be overestimating the chloride-surface interaction and, consequently, yielding an artificial increase in core binding energies, due to the fact that waters do not relax when the ion is introduced.

It may also be that our configurations are not properly representing the local environment of the chloride ion, as probed by the spin-orbit splitting of the 2p, though here the current experimental and theoretical data, in our view, do not allow for any definitive conclusions. On the one have,~\citeauthor{kong2017coexistence}~\cite{kong2017coexistence} obtained an experimental difference between the L$_2$ and L$_3$ edges of 2.1 eV. On the other hand, our calculated splitting for chloride--ice is of roughly 1.7 eV, a value consistent with splitting between L$_2$ and L$_3$ edges in the NaCl crystal (see table~\ref{tab:ham_bas}) and slightly larger than the roughly 1.6 eV we obtained for the gas-phase ion, our droplet results and the experimental work of~\citeauthor{Pelimanni2021}~\cite{Pelimanni2021} find for the solvated ion. It is also interesting to note that the 2.1 eV splitting is much larger than the 1.3 eV splitting observed by~\citeauthor{parent2011hcl}~\cite{parent2011hcl}, that is closer to our results. 

In the case of the K edge, a problem with adequate sampling could be the reason for our large overestimation of the experimental binding energies, since from figure~\ref{fig:eom-cl-hcl-average300-210} we see that K edge energies are extremely sensitive to the configuration. At this stage we lack a better classical polarizable force field that can represent both the water-chloride and water-water interactions in the ice QLL. Due to this, the question of whether (and if so, how) better sampling would affect the K edge remains an open question.

\begin{figure*}[htp!]
\centering
\includegraphics[width=0.95\linewidth]{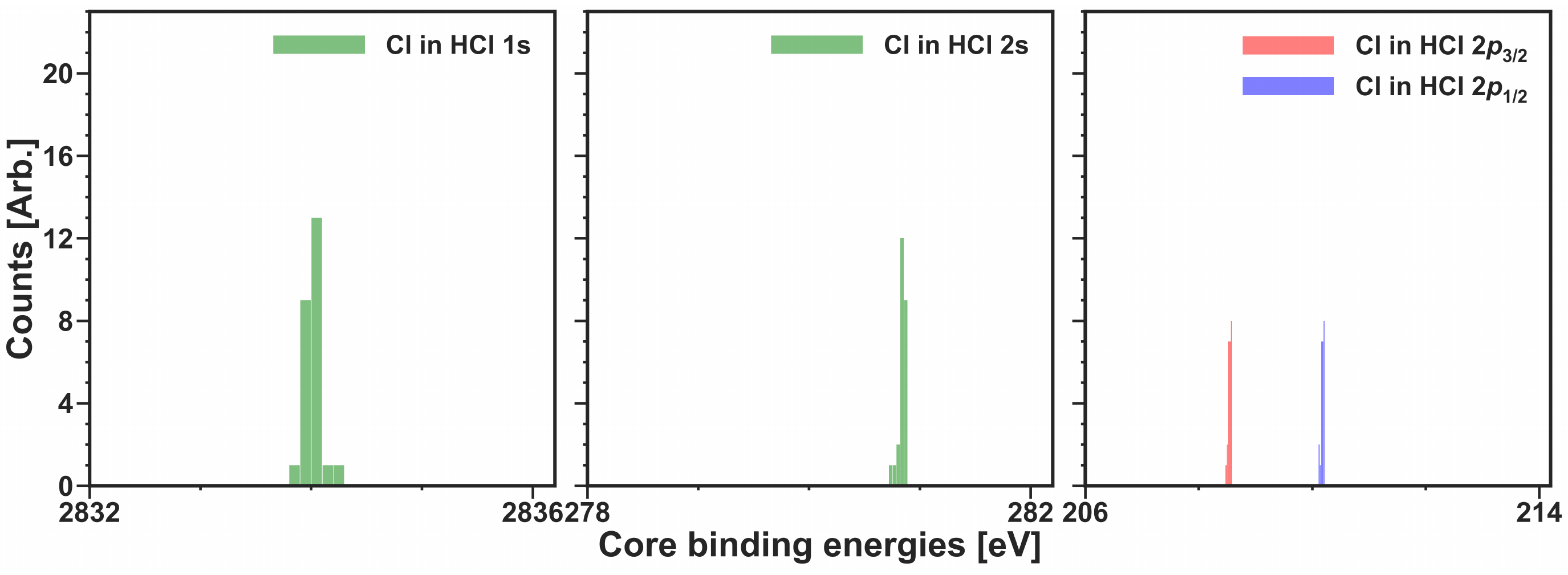}
\includegraphics[width=0.95\linewidth]{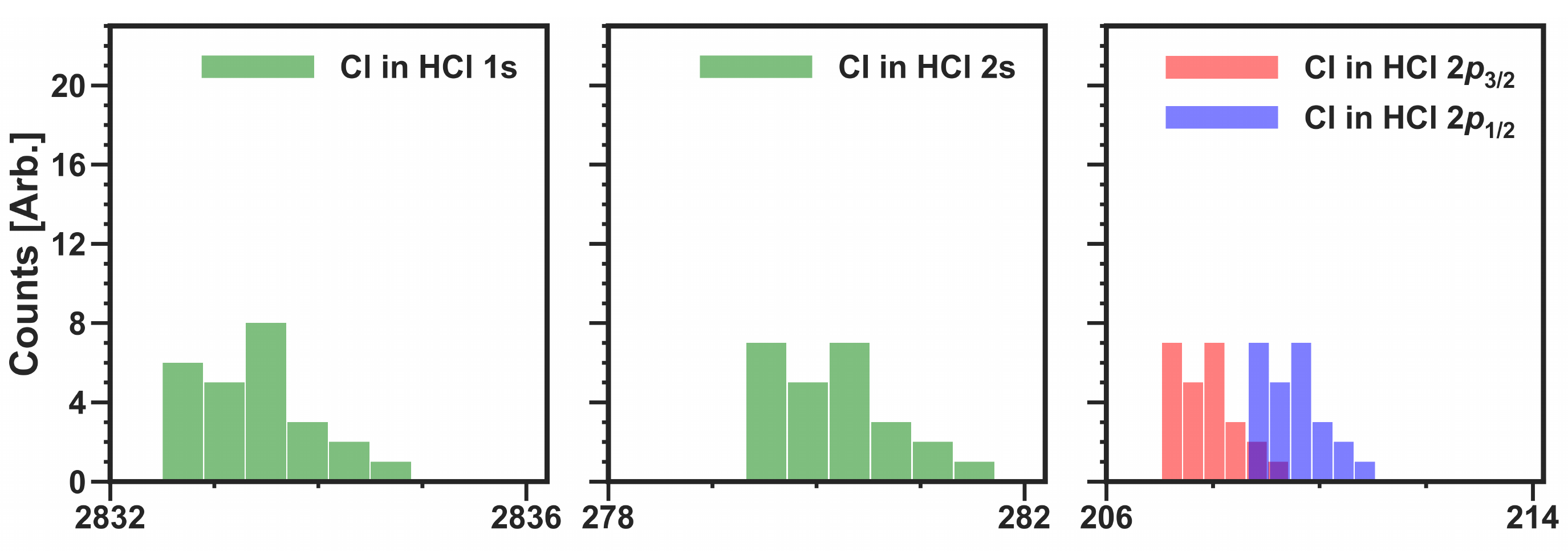}
\includegraphics[width=0.95\linewidth]{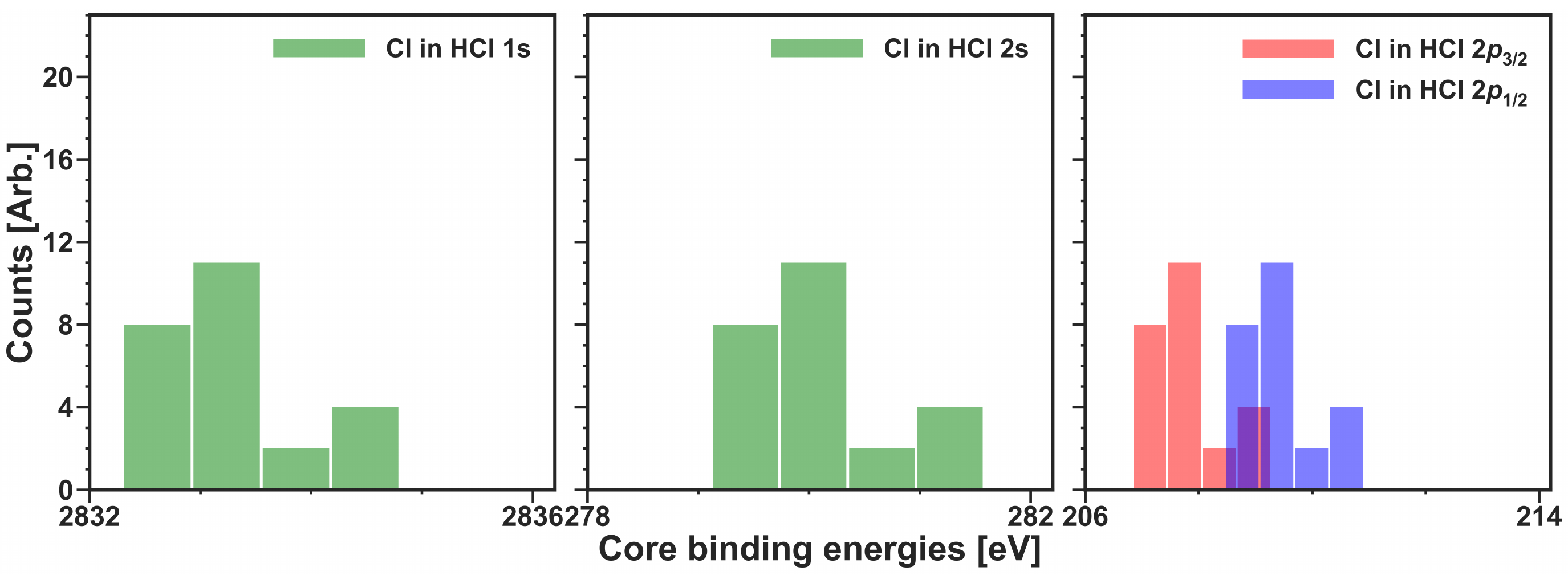}
\caption{\label{fig:eom_hcl-average_rigid} CC-in-DFT chlorine K, L$_1$, L$_2$ and L$_3$ triple-zeta binding energies for HCl adsorbed on ice surfaces at 210 K (results averaged over 25 snapshots) employing as structural models (A) the original CMD snapshots (top); (B) reoptimizing the HCl molecule while constraining the ice surface to retain the atomic positions of model A (middle); and (C) reoptimizing the HCl and four nearest water molecules, while constraining the rest of the ice surface to retain the atomic positions of model A (bottom)}.
\end{figure*}

Unlike the two chloride systems discussed above, for HCl a straightforward use of the molecular dynamics simulations snapshot yields results which are essentially the same as those for the isolated molecule.  This indicates that in these snapshots there are, in effect, all but residual interactions between HCl and the ice ($\Delta^g$BE(A) values are very small and around 0.15 eV for all edges). By inspecting the top of figure~\ref{fig:eom_hcl-average_rigid}, this becomes quite clear: in spite of averaging over 25 snapshots, there is essentially no spread in binding energy values, which would otherwise be the case if there were stronger interactions with the surface. This is consistent with the findings of~\citeauthor{Woittequand2007}~\cite{Woittequand2007} that the adsorption energy of HCl on ice (of the order -0.2 eV) is quite small in absolute value. 

If we take the snapshots as starting point for geometry optimizations of the HCl molecule, keeping the ice structure constrained to the original CMD configurations, we see small but non-negligible change in the binding energies (BE(B)) for the different edges, so that now instead of the slight increase of binding energies seen at first, we start to see a move towards lower binding energies (($\Delta^g$BE(B) of about -0.50 eV), that is, towards the experimental trend (HCl on ice binding energies being lower than gas-phase ones).  Similarly to the chloride on ice system, there is a large spread in values (of around 2-3 eV for the K and L edges, see middle of figure~\ref{fig:eom_hcl-average_rigid}).

Upon obtaining configurations in which we also optimize the waters nearest to the HCl molecule, we observe a further decrease in binding energies (BE(C)) that, though in the direction of experiment, is too small to bring our calculations to the same agreement with experiment as seen for the L edge of the chloride--ice system discussed above, and we see discrepancies of around 5.7-6 eV. The discrepancy between theory and experiment for HCl--ice K edge binding energies are also around 3 eV larger than for the chloride--ice system.
 
One possible issue with these simulations on ice is that, at 210K, the  temperature of the simulations is somewhat lower than that of the experiment. This makes it worthwhile to explore the effect of the temperature on the ice structures, and see to which extend the changes would affect the binding energies.

To this end, we have carried out additional classical MD simulations for HCl under the same conditions as done for 210K, one at 235K and another at 250K, and selected a single snapshot of each to carry out exploratory CC-in-DFT calculations. In figures S1 and S2 we show respectively the structures for HCl and Cl$^-$ at the two temperatures. 

From these figures, and keeping in mind the structures at 210K shown in figure~\ref{fig:modelstofollow}, we see the progressive desorganization of the upper layers of the interface when passing from 210K to 235K (though the innermost two layers remain rather well structured), and a fairly substantial loss of structure going from 235K to 250K.

In spite of these significant changes in structure, between the different temperatures there are little changes to the binding energies, as can be seen from table S5 in the supplementary information. Even though we have only one structure, and therefore we cannot strictly speaking compare to the averaged results for 210K, the results in table S5 suggest nevertheless that temperature effects cannot play a major role in modifying the binding energies if the HCl molecule remains essentially bound (and with an internuclear distance not far from the gas-phase value), as we discuss in the following.

\subsection{A closer look on the HCl-water interaction}

\begin{figure*}[htp!]
\centering
\includegraphics[width = 16.0cm, trim = 0cm 0cm 0cm 0cm]{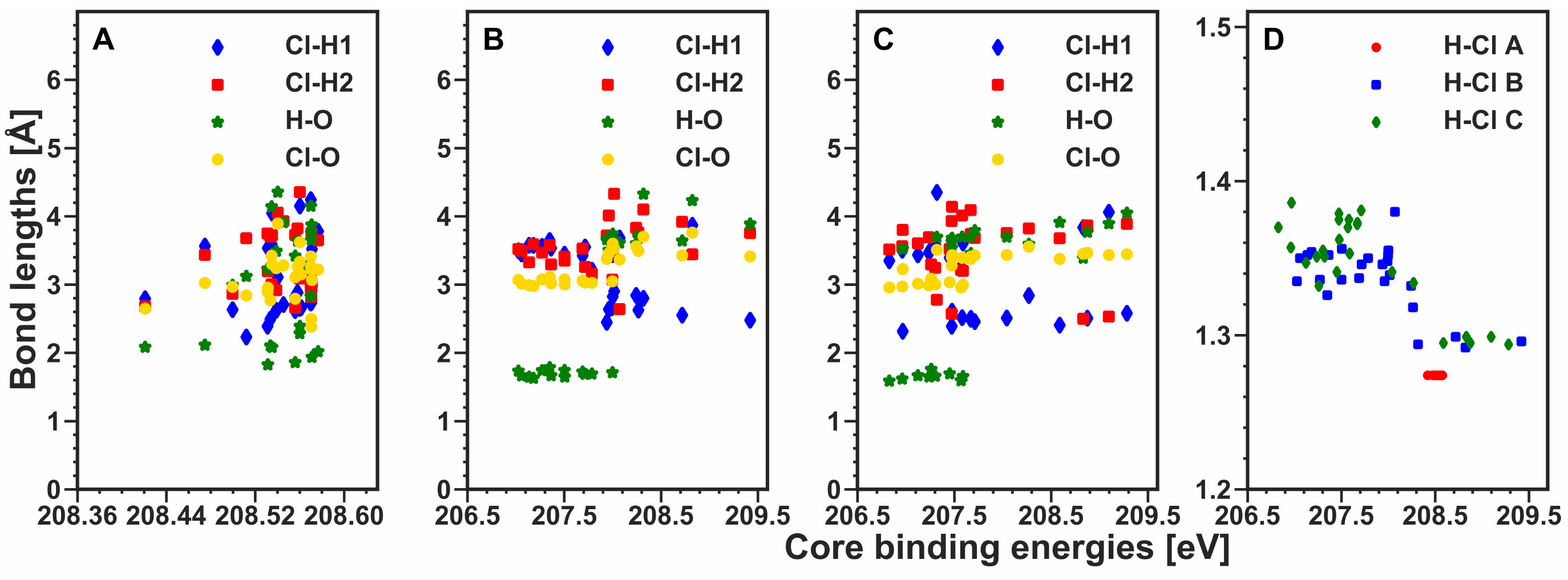}
\caption{\label{fig:eom-be_atoms} Scalar ZORA chlorine 2p binding energies as a function of HCl-H$_{2}$O intra- and inter-molecular distances of HCl adsorbed on ice surfaces at 210 K for the 25 snapshots, employing as structural models (A) the original CMD snapshots; (B) reoptimizing the HCl molecule while constraining the ice surface to retain the atomic positions of model A; and (C) reoptimizing the HCl and four nearest water molecules, while constraining the rest of the ice surface to retain the atomic positions of model A. In panel D the BEs with respect to the HCl bond lengths in models A, B and C are shown.}
\end{figure*}

The contrast between chloride and HCl results, and the changes (albeit modest) in binding energies observed for HCl depending on the strucural model for the HCl-ice surface interaction sites discussed above, call for a closer look at how the structural parameters affect the calculated binding energies, as shown in figure~\ref{fig:eom-be_atoms}.  Considering first the HCl internuclear distance (panel D), we see that for the original snapshots from CMD simulations of~\citeauthor{woittequand2007classical}~\cite{woittequand2007classical} (model A) one obtains  essentially the same binding energies which, as discussed above, are nearly indinstiguishable from the gas-phase ones. 

Upon optimizing the HCl position while keeping the surface unchanged (model B), we see a significant change in that internuclear distances increase for all snapshots with respect to model A, to values between 1.28 and 1.38 \AA; furthermore, we can identify three categories of points: those for internuclear distances around  1.28 \AA, which are associated to larger core binding energies (right of the figure), those for internuclear distances between 1.32 and 1.36 \AA, which are associated with lower core binding energies (left of the figure), and the third cluster for internuclear distances between 1.28 and 1.34 \AA, but which exhibit roughly the core same binding energies (around 208 eV). 

The optimization of the HCl and nearest water molecules (model C) accentuates somewhat the trend of increased internuclear distances in the region for lower core binding energies seen for model B. We observe more snapshots with internuclear distances larger than 1.36 \AA, which are 1 to 2 eV lower than the core binding energies for model A (and we note that, contrary to the gas-phase results, relatively small changes in internuclear distance produce a significant shift of core binding energies). However, the average core binding energy only shows modest changes with respect to model A due to the fact that there remain several structures with core binding energies larger than 208.5 eV.  

Apart from the H-Cl distance, we see significant changes in the distances between the hydrogen in HCl and the nearest oxygen atoms of the surface: while for model A, the large variation in this O-H distance does not significantly affect the binding energies, for models B and C we can distinguish two types of distances: longer ones (around 3.5 \AA) for which the core binding energies are generally above 208 eV, and shorter ones (around 1.6 \AA) associated with lower core binding energies.

\begin{figure}[htp!]
\centering
\begin{minipage}{\linewidth}
\includegraphics[width=\linewidth]{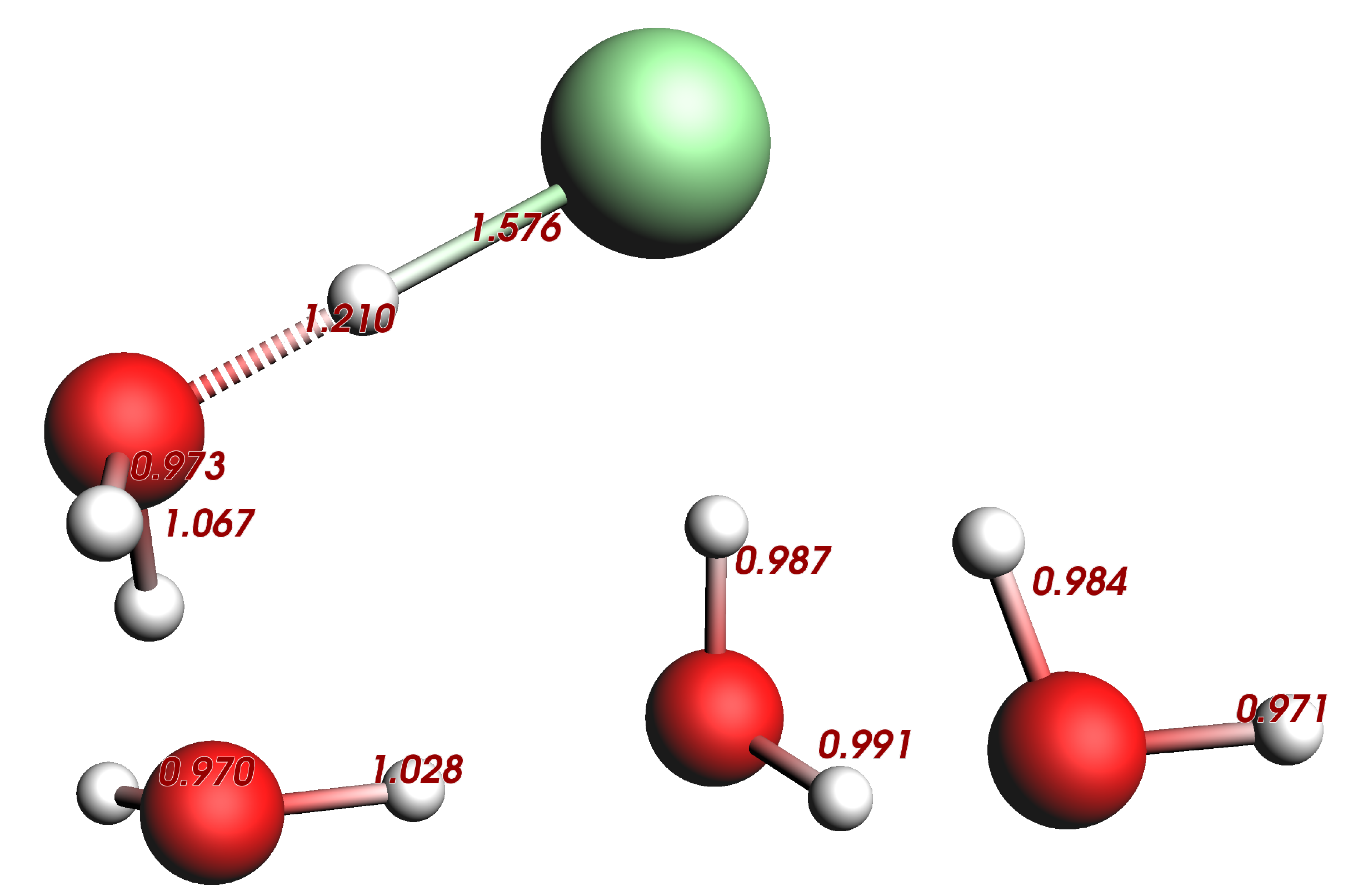}
\end{minipage}
\begin{minipage}{\linewidth}
\includegraphics[width=\linewidth]{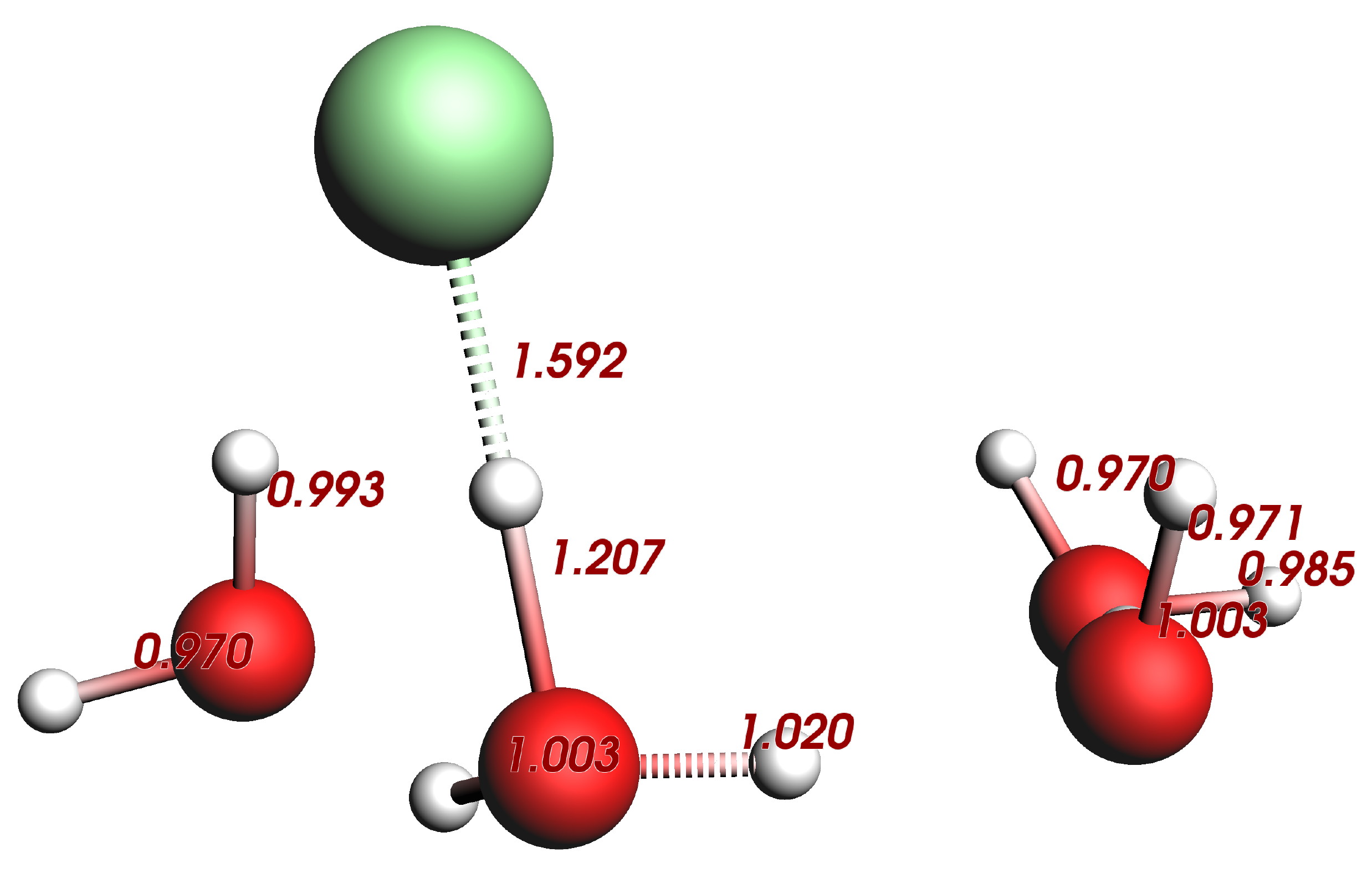}
\end{minipage}
\caption{\label{fig:microsolvated-view} Structures for a microsolvated HCl molecule originating from a snapshots of model C, with nearest 4 waters shown. Internuclear distances (in \AA) are indicated in the figure. Binding energies (in eV) for the [K, L$_1$, L$_2$, L$_3$] edges, obtained with CC-in-DFT (for the microsolvated system) are respectively [2832.6, 279.3, 208.7, 207.1] (Top) and [2831.9, 278.7, 208.0, 206.4] (Bottom).}
\end{figure}

Taken together, these observations suggest that the lowewring of the core binding energies is closely connected to the concerted increase in the HCl internuclear distance and decrease of the oxygen surface atoms and the hydrogen of HCl, which would represent the initial stage of the (pre)dissociation of HCl mediated by the surface. Two such configurations can be more clearly visualized in figure~\ref{fig:microsolvated-view}, which depicts the spatial arrangement of HCl and its nearest four water molecules for two situations in which structural relaxation is taken into account. The figure also contains the core binding energies obtained for each microsolvated cluster. We observe that, already in such simplified models, binding energies are pretty sensitive to relatively small changes in structure. One can also identify in the figure cooperative effects coming from the elongation of certain O-H bonds in the water molecules as the HCl molecule gets closer (with the elongation of the H-Cl bond and the interaction of the hydrogen of HCl and the oxygen of the nearest water). A further investigation of the influence of predissociation of HCl would require more extensive CMD simulations for the surface, which are beyond the scope of this work.
 
To further explore this point, we have considered two additional models: (a) one in which a +1 point charge is placed at a given distance $r$ from the chloride ion; and (b) one in which we employ one snapshot of the chloride droplet model to construct a HCl droplet model, by placing the added hydrogen atom near the chloride or a given oxygen (to simulate the H$_3$O$^+$ species), and performing a constrained optimization (fixing all atoms but the hydrogens belonging to the same species the hydrogen has been attached to). The results for these models are found in table~\ref{tab:binding-energies-changes-hydronium} and table S6 of the supplementary material.

\begin{table}
\caption{\label{tab:binding-energies-changes-hydronium} Core binding energy shift ($\Delta_{BE}$, in eV) between Cl$^{-}$ and the HCl systems as a function of the distance $r$ (in~\AA) between the chlorine and the hydrogen atom, for (a) the electrostatic gas-phase model based on the pair of charges (chloride and a +1 point charge representing the hydrogen); and (b) the [HCl(H$_2$O)$_{50}$] droplet model structure based on the [Cl(H$_2$O)$_{50}$]$^-$ droplet model discussed above, in which the additional hydrogen is found near the different heavy centers (chlorine, oxygens). See text and supplementary information for further details and approximate core electron binding energies.} 
\centering
\begin{tabular}{r rrr p{0.1cm} rrr}
\hline
\hline
%
& \multicolumn{7}{c}{Core Binding Energies} \\
\cline{2-8}
& \multicolumn{3}{c}{droplet model} && \multicolumn{3}{c}{gas-phase model} \\
\cline{2-4}
\cline{6-8}
$r$     &    K       &       L$_1$   &       L$_2$/L$_3$ && K        &       L$_1$   &       L$_2$/L$_3$\\
\hline
  1.306   &       7.48    &       7.53    &       7.69    &&      8.59    &       8.73    &       8.80    \\
  2.559   &       3.05    &       3.06    &       3.05    &&      5.70    &       5.71    &       5.75    \\
  3.489   &       2.81    &       2.81    &       2.82    &&      4.16    &       4.17    &       4.18    \\
  5.690   &       2.32    &       2.33    &       2.33    &&      2.53    &       2.53    &       2.54    \\
  10.256  &       1.60    &       1.64    &       1.64    &&      1.40    &       1.40    &       1.40    \\

\hline
\hline
\end{tabular}
\end{table}

Starting with the gas-phase model (a), we observe that at distances slightly larger than the gas-phase equilibrium ($r = 1.306$~\AA), the $\Delta_{BE}$ values are rather close to the values for the molecular HCl system (8.6-8.8 eV vs roughly 10 eV in table~\ref{tab:ham_bas}). As $r$ is increased to $r = 2.559$~\AA, distance already much larger than those sampled by our MD simulations and shown in figures~\ref{fig:eom-be_atoms} and~\ref{fig:microsolvated-view}, there's a significant drop in $\Delta_{BE}$ to 5.7 eV, and then a relatively smooth decrease for larger distances.  Interestingly, at around $r = 5.690$~\AA, $\Delta_{BE}$ drops to around 2.5 eV, which is in the order of magnitude of the experimental chemical shift reported by~\citeauthor{kong2017coexistence}~\cite{kong2017coexistence} (2.2 eV), and also close to the value by~\citeauthor{parent2011hcl}~\cite{parent2011hcl} (2.6 eV). While not shown, we have investigated how far the +1 point charge would have to be in order for us to obtain the gas-phase Cl$^-$ value, and at $r = 100$~\AA~there are still small differences (around 0.1 eV). 

For model (b) we have the same qualitative trend, but with an interesting difference: while the $\Delta_{BE}$ value at $r = 1.306$~\AA~is still relatively large (around 7.5 eV), comparing it with the gas-phase $\Delta_{BE}$ value shown in~\ref{tab:ham_bas}, we can infer an environment effect of around 2.5 eV--which is already much larger than the effects shown in table~\ref{tab:ccsd-ice-hcl}, and in line with figures~\ref{fig:eom-be_atoms} and~\ref{fig:microsolvated-view}. For $r = 2.559$~\AA, due to the effect of the screening by the water molecules in the droplet (see figure S3 in the supplementary information), the $\Delta_{BE}$ value (arount 3 eV) is nearly 3 eV lower than the one from the gas-phase. Likewise, at $r = 3.489$~\AA~$\Delta_{BE}$ is already below 3 eV (and a little over 1 eV smaller than the gas-phase value). 

For larger $r$ values model (b) follows roughly the behavior of model (a), which may be due to the small size of the droplet. Contrary to the gas-phase model, our simulations do not allow us to probe much longer distances than 10~\AA, but the relative small difference to the anionic systems suggests in our view that the anionic systems would be indeed rather good models for diluted solutions. In subsequent work it will be interesting to investigate how screening will alter the rate of convergence towards the anionic result, now that polarizable force fields of similar accuracy to those employed here and that can handle counter-ions are starting to become available~\cite{vallet2021nacl}. 

Whatever the case, these results suggest that for a solvated system, chemical shifts compatible to those observed in experiment occur for values of $r$ between 4 and 6~\AA. 

While these results only consider single structures, and therefore must be viewed as providing semi-quantitative evidence, they help to understand that the poor agreement with experiment for the simulations of HCl on ice shown in table~\ref{tab:ccsd-ice-hcl} likely comes from an inadequate structural model, which assumes and retains a bound (molecular) picture for HCl, when it would seeem that a more suitable situation resembles the existance of an ion pair, or some other intermediate situation. Unfortunately, we do not currently possess the adequate tools to further explore this problem, and in any case such a undertaking requires a dedicated study.

\section{Conclusion\label{sec:conclusion}}

In this article we report the application of the relativistic CVS-EOM-IP-CCSD-in-DFT approach to obtain core electron binding energies for chlorinated species (Cl$^-$ and HCl) for the air-ice interface which is of great interest for atmospheric chemistry and physics, as well as of Cl$^-$ in aqueous solution (which allows us to differentiate isotropic and non isotropic solvation of the anion).  In our coupled cluster calculations, we employ the molecular mean-field Dirac-Coulomb-Gaunt Hamiltonian, which accurately accounts for spin-same orbit and spin-orther orbit interactions.

These calculations are based upon structural models considering, for both droplets and ice surfaces, the halides and the nearest 50 and 200 water molecules respectively. Based on these, embedding models in which all water molecules were treated at DFT level while the halide species were trated with coupled cluster have been assessed, and their relative accuracy verified against reference DFT calculations on the whole system. There, we have found that subsystem DFT calculations, in which both the halide and 50 nearest water molecules in the environment were relaxed in the presence of each other, were well-suited for both the neutral (HCl) and the charged (Cl$^-$) subsystems, showing small and systematic errors for all edges.

The accuracy of our protocol has been shown for the water droplet case, for which we obtain L$_2$ and L$_3$ CBS-corrected core binding energies in very good agreement but overestimate the most recent experiments by~\citeauthor{Pelimanni2021}~\cite{Pelimanni2021} in solution by around 1 eV for each edge, while obtaining nearly the same energy splitting (1.66 eV, due to spin-orbit coupling) as experiment (1.6 eV) between the L$_2$ and L$_3$ edges. We consider that remaining discrepancies are likely due, in part, to the different temperatures in which theoretical and experiment results have been obtained, and also due to the lack of corrections for higher excitations in our calculations. 

Our theoretical values for the L edges of chloride on ice surfaces show systematic differences to the droplet model (binding energies roughly 1.7 eV smaller). They are also in fairly good agreement with experiment for the L$_2$ and L$_3$ edges, though with larger discrepancies than for water droplets, with theoretical values overestimating experiment by 1.8 and 2.3 eV respectively. 

While that may partially arise from the shortcomings of our protocol to obtain structures for chloride (classical MD simulations on HCl followed by constrained geometry optimization of the chloride position, due to the lack of suitable force fields), it should be pointed out that our theoretical spin-orbit splitting between the L$_2$ and L$_3$ edges is consistent (1.56 eV) with that in the droplet model, with experiments in solution, and in the NaCl crystal. The experimental splitting on ice reported by~\citeauthor{kong2017coexistence}~\cite{kong2017coexistence}, on the other hand, is somewhat larger (2.1 eV), and merits to be further investigated from both a theoretical and experimental point of view.

For the K edge we observe that chloride binding energies on ice are 1.1 eV smaller than those in the droplet model, qualitatively in line with the results for the L edges and with the overall experimental trend of lowering the core binding energies when going from a solution to a surface for chloride. However, theory strongly overestimates (by 13.4 eV) the experimental K edge binding energies.  We cannot at this stage pinpoint the factor(s) driving this discrepancy, though we note that in contrast to the droplet model, there are very significant variations in the K edge binding energies depending on the snapshot used. In our view, this calls for additional simulations once a better force field for the MD simulations is available, in order to confirm whether or not configuration sampling is a significant source of bias to the theoretical binding energies. However, at this point we are not equipped to carry out such simulations.

We also report estimates for the valence binding energy of water in the ice surfaces, which are in good agreement with experiments at low temperatures. For that we have followed the procedure previously employed for obtaining the water valence band for water in droplets, in which the use of the SAOP model potential to describe the environment yields valence binding energies as a by-product of the setup of the embedded coupled cluster calculations at no additional cost. 

In contrast to the chloride results, our results for molecular HCl on ice surfaces show poor quantitative agreement with experiment. We obtain core binding energies that are significantly higher than experimental ones (nearly 6 eV for the L edges, and nearly 17 eV for the K edge), due to the fact that there are almost negligible environmental effects and, therefore, the results are essentially the same as those for gas-phase HCl.  

We have assessed to a limited extent the importance of temperature effects on the binding energies, by comparing the CC-in-DFT core binding energies based on single structures taken from classical MD simulations of HCl at higher temperatures (235K and 250K) to those averaged over 25 snapshots at 210K. We found that in spite of the important desorganization of the ice stucture as the temperature increases, there are no significant changes in binding energies for Cl in HCl.

From an analysis of small microsolvated clusters, and of two models to capture the variation of the core binding energies as a function of the H-Cl distance  (a simple gas-phase one, and droplet models for HCl constructed from the chloride droplet model shown to be reliable), we were able to trace the root cause of these lack of sensitivity to the environment to the inability of our structural models to account for the (pre)dissociation of HCl molecule upon interaction with water molecules around it, resulting in the transfer of the proton to the first and second solvation layers. 

In spite of their limitations, and notably the lack of any configurational averaging, the gas-phase electrostatic and HCl droplet models provide evidence that, in order to be consistent with the experimentally observed chemical shifts between chloride and HCl of about 2.2-2.6 eV, the proton in the latter should likely be distant from the chlorine atom by around 5-6~\AA, a situation which is compatible with it participating in the hydrogen bond network. As such, a perhaps more suitable description for HCl on ice is that of an ion pair rather than of a molecular species. In order to properly characterize this system it appears to us that a central ingredient would be a MD approach that could (at least) approximately account for such ion-pair formation, and ideally describe the exchange of hydrogens between different water molecules, but such studies fall outside the scope of this work.


\section*{Acknowledgements}

We acknowledge support by the French government through the Program “Investissement d'avenir” through the Labex CaPPA (contract ANR-11-LABX-0005-01), the I-SITE ULNE project OVERSEE and MESONM International Associated Laboratory (LAI)  (contract ANR-16-IDEX-0004), the Franco-German project CompRIXS (Agence nationale de la recherche ANR-19-CE29-0019, Deutsche Forschungsgemeinschaft JA 2329/6-1), CPER CLIMIBIO (European Regional Development Fund, Hauts de France council, French Ministry of Higher Education and Research) and French national supercomputing facilities (grants DARI A0070801859, A0090801859 and A0110801859).

\bibliography{reference,ms} 
\bibliographystyle{achemso}

\end{document}



\title{Supplemental information -- Simulating core electron binding energies of halogenated species adsorbed on ice surfaces and in solution with relativistic quantum embedding calculations}

\author{Richard A. Opoku}
\author{Céline Toubin}

\author{André Severo Pereira Gomes}
\email{andre.gomes@univ-lille.fr}
\affiliation{Univ.\ Lille, CNRS,
UMR 8523 -- PhLAM -- Physique des Lasers Atomes et Mol\'ecules, F-59000 Lille, France}

\date{\today}

\maketitle


\section{Additional computational details}

As indicated in the manuscript, our base models contain a total of 200 water molecules. From them, we investigated two directions: first, at DFT-in-DFT level, we considered a smaller model, in which the 50 waters closest to the halogen were retained, and for a single snapshot, we have investigated the effect of relaxing the density of the $n$ ($n = 0, 10, 20, 30, 40$ for \textbf{EM1} and \textbf{EM2}, and also $n = 50$ for \textbf{EM1}) water molecules closest to the active subsystem, via freeze-thaw iterations, on the DFT and DFT-in-DFT orbital energies. If orbital energies provide a poor model for comparing to experiment due to the lack of orbital relaxation, they are well-defined quantities, are obtainable for all model sizes and provide a qualitatively correct picture of the changes in electrostatic interaction between the surface and the halogens as the number of water molecules is increased.

Second, we have investigated, also for a single snapshot, the effect of the number of water molecules on the BEs for systems containing 50, 100, 150 and 200 water molecules for model \textbf{EM1}. Due to constraints in our computational resources, and also in the perspective of performing both DFT-in-DFT and CC-in-DFT calculations, we have employed a variant of model \textbf{EM2} (denoted by \textbf{EM2$^{V}$}), in which only the nearest water molecule to the halogen is added to the active subsystem. In the CC-in-DFT calculations associated with such tests, we have restricted the virtual spinor space to include only those with energies up to 100 Hartree, and therefore exclude high-lying virtual spinors which are very important for obtaining accurate EOM-CC BEs. As such, our results here should be viewed as semiquantitative at best.

From these two investigations, \textbf{we arrived at the final models for which we carried out the conformational averaging} by considering snapshots from CMD simulations: for both ice and water droplet, these consisted of the halogen species and a total of \textbf{50 water molecules.}

\clearpage

\section{Hamiltonian, basis set and molecular properties convergence}

\begin{table*}[htp]
\caption{\label{tab:ham_bas} CVS-EOM-IP-CCSD chlorine core binding energies (in eV) for HCl and Cl$^{-}$ in gas-phase for Dirac-Coulomb based Hamiltonians ($^4$DC and $^2$DC$^M$) and employing triple-zeta basis sets as well as values extrapolated to the complete basis set limit (CBS). In addition to those, we present core binding energies obtained via the analogue of Koopmans theorem for DFT~\cite{Chong2002}, employing the SAOP model potential for the ZORA Hamiltonian. In parenthesis we presente the differences with respect to the CVS-EOM-IP-CCSD $^2$DCG$^M$ results with triple-zeta basis sets, which we take as reference ($^2$DCG$^M$ are found in the body of the manuscript). Apart from the energies for the individual edges, we provide the core binding energy shift ($\Delta_{BE}$, in eV) between HCl and Cl$^{-}$.}
\begin{tabular}{l l l r r r r}
\hline
\hline
Species & Hamiltonian & Method & K & L$_1$ & L$_2$  &L$_3$ \\
\hline
HCl      & $^4$DC      & EOM    & 2835.79 (  1.89) & 280.77 (  0.08) & 210.18 (  0.16) & 208.48 (  0.09) \\
         & $^2$DC$^M$  & EOM    & 2835.78 (  1.88) & 280.77 (  0.08) & 210.18 (  0.16) & 208.48 (  0.09) \\
         & & & & & \\
Cl$^{-}$ & $^4$DC      & EOM    & 2826.06 (  1.89) & 270.81 (  0.08) & 200.26 (  0.16) & 198.56 (  0.09) \\
         & $^2$DC$^M$  & EOM    & 2826.05 (  1.89) & 270.81 (  0.08) & 200.26 (  0.16) & 198.56 (  0.09) \\
         & & & & & \\
$\Delta_{BE}$ & $^4$DC      & EOM    &    9.73          &   9.96          &   9.92          &   9.92          \\
         & $^2$DC$^M$  & EOM    &    9.73          &   9.96          &   9.92          &   9.92          \\
\hline
\hline
\end{tabular}
\end{table*}

\begin{table*}
\caption{\label{tab:ccsd-ice-hcl_basis_sets_dz_tz_qz} Basis set convergence for CVS-EOM-IP-CCSD-in-DFT for the embedded system on ice. The values in the table for each basis set correspond to the binding energies for each edge, averaged over 25 snapshots. In the case of HCl, the structural model used is the one in which both HCl and nearest water neighbors are optimized (model C).  Apart from the energies for the individual edges, we provide the core binding energy shift ($\Delta_{BE}$, in eV) between HCl and Cl$^{-}$.}
\begin{tabular}{lll rrrrr r}
\hline
\hline
System   & Environment &Edge&  BE(DZ) & BE(TZ)  & BE(QZ)   & BE(CBS) & Diff(CBS-TZ) \\
\hline
 HCl     & ice     & K      &  2833.73 & 2833.40 &  2833.94 & 2834.33 & 0.93    \\
         &         & L$_1$  &   280.09 &  280.15 &   280.75 &  281.19 & 1.04    \\
         &         & L$_2$  &   209.52 &  209.49 &   210.13 &  210.60 & 1.11   \\
         &         & L$_3$  &   207.90 &  207.86 &   208.54 &  209.04 & 1.18   \\
&&&&&&&&\\
Cl$^{-}$ &         & K      & 2825.89 & 2827.70 &  2828.79 & 2829.59 & 1.89    \\
         &         & L$_1$  &  271.86 &  274.90 &   275.07 &  275.19 & 0.29   \\
         &         & L$_2$  &  201.31 &  204.41 &   204.47 &  204.51 & 0.10   \\
         &         & L$_3$  &  199.68 &  202.70 &   202.91 &  203.06 & 0.36  \\
&&&&&&&&\\
$\Delta_{BE}$	&&	 K      	&	7.84	&	5.70	&	5.15	&	4.74	&	-0.96	\\
	&&	 L$_1$  	&	8.23	&	5.25	&	5.68	&	6.00	&	0.75	\\
	&&	 L$_2$  	&	8.21	&	5.08	&	5.66	&	6.09	&	1.01	\\
	&&	 L$_3$  	&	8.22	&	5.16	&	5.63	&	5.98	&	0.82	\\
\hline
\hline
\end{tabular}
\end{table*}

\begin{table*}
\caption{\label{tab:hcl-bond-be} Variation of $^2$DCG$^M$ chlorine core binding energies (in eV) with respect to the $r_\text{H-Cl}$ distance (in \AA) for gas-phase HCl. Results were obtained with triple-zeta basis sets. CCSD energies (E$_\text{CCSD}$, in atomic units) for the ground state are also shown. For the L$_2$ and L$_3$ we present the weighted averaged value.}
\begin{tabular}{lllll}
\hline
\hline
                &             & \multicolumn{3}{c}{Core Binding Energies} \\
\cline{3-5}
$r_\text{H-Cl}$ & E$_\text{CCSD}$  & K & L$_1$ & L$_2$/L$_3$ \\
\hline
1.074&-462.027&2833.66&280.72&209.22 \\
1.124&-462.045&2833.74&280.72&209.22 \\
1.174&-462.055&2833.80&280.71&209.22 \\
1.224&-462.060&2833.86&280.70&209.22 \\
1.274&-462.062&2833.90&280.69&209.21 \\
1.324&-462.060&2833.94&280.68&209.19 \\
1.374&-462.057&2833.96&280.65&209.17 \\
1.424&-462.051&2833.98&280.63&209.15 \\
1.474&-462.044&2834.00&280.60&209.13 \\
1.524&-462.034&2834.00&280.58&209.10 \\
\hline
\hline
\end{tabular}
\end{table*}
%


\begin{table*}[htp]
\caption{\label{tab:dipole-moment-hcl} Variation with respect to the number of water molecules in the environment ($n$ H$_2$O) of the permanent dipole moments ($\mu_0$, in Debye) obtained from the supermolecular DFT calculations on the ice subsystem without the halogenated species, and from DFT-in-DFT scalar ZORA calculations for the HCl system, employing the SAOP model potential. The structures considered are those employed to obtain the CC-in-DFT binding energies shown in figure 3 in the manuscript. For the DFT-in-DFT calculations, The dipole moments are broken down into the contributions from each subsystem and the total value (the latter calculated as  $\mu^\text{total}_0 = (\sum_{i=x,y,z} [\mu^\text{ice}_i + \mu^\text{active}_i])^{1/2}$. The permanent dipole moment of gas-phase HCl is 1.08 D.}
\begin{tabular}{lr p{0.5cm} r p{0.5cm} rrr}
\hline
\hline
                  &        &&   DFT  && \multicolumn{3}{c}{DFT-in-DFT} \\
\cline{4-4}
\cline{6-8}
active subsystem  & $n$    &&   ice  &&  ice  & active & total \\
\hline
HCl               &   0    &&  ---   &&  ---  &  1.02  &  1.02 \\
                  &   8    &&  5.83  &&  5.85 &  1.46  &  6.46 \\ 
                  &  50    && 14.88  && 15.39 &  1.56  & 16.54 \\
                  & 100    && 14.88  && 15.44 &  1.57  & 16.62 \\
                  & 150    && 14.88  && 15.05 &  1.56  & 16.21 \\
                  & 200    && 14.88  && 15.22 &  1.57  & 16.40 \\
&&&&&&&\\
HCl-H$_2$O        &   0    &&  ---   && ---   &  2.66  &  2.66 \\
                  &   7    &&  5.71  &&  5.71 &  3.75  &  6.31 \\ 
                  &  49    && 12.11  && 12.79 &  3.92  & 16.37 \\
                  &  99    && 12.11  && 12.77 &  3.93  & 16.41 \\
                  & 149    && 12.11  && 12.36 &  3.89  & 15.96 \\
                  & 199    && 12.11  && 12.54 &  3.90  & 16.15 \\
\hline
\hline
\end{tabular}
\end{table*}


\clearpage

\section{Effects of temperature on binding energies}

To investigate the effects of temperature on the halogen-ice systems, we have carried out additional CMD simulations for the HCl-ice system at 235K and 250K, employing the same non-polarizable force field used for calculations at 210K. Apart from the change in temperature, the CMD calculations were carried out in the same conditions as for the 210K ones.

\begin{figure*}[htp!]
\centering
\begin{minipage}{0.67\textwidth}
\includegraphics[width=\linewidth]{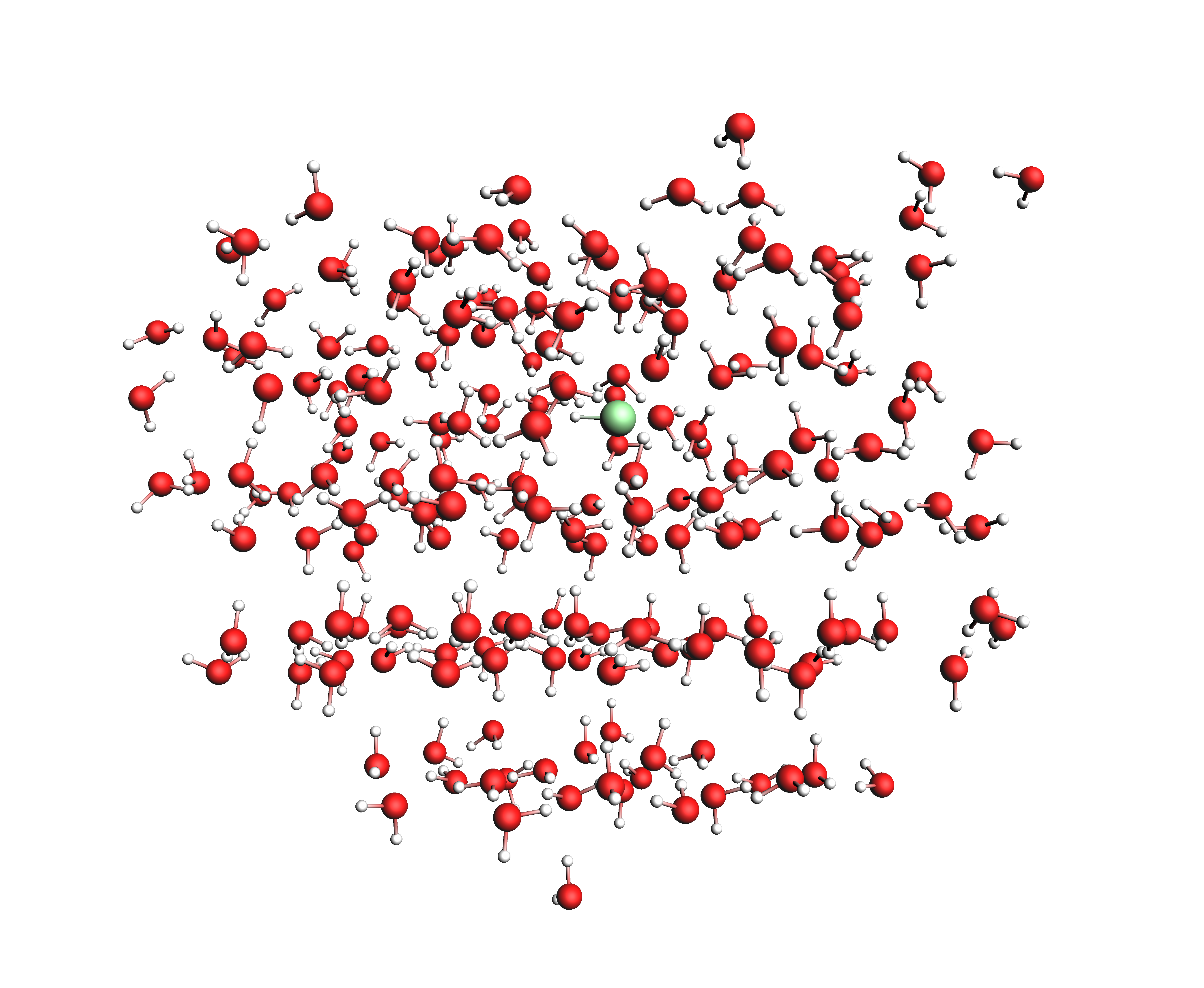}
\caption*{235 K}
\end{minipage}
\begin{minipage}{0.67\textwidth}
\includegraphics[width=\linewidth]{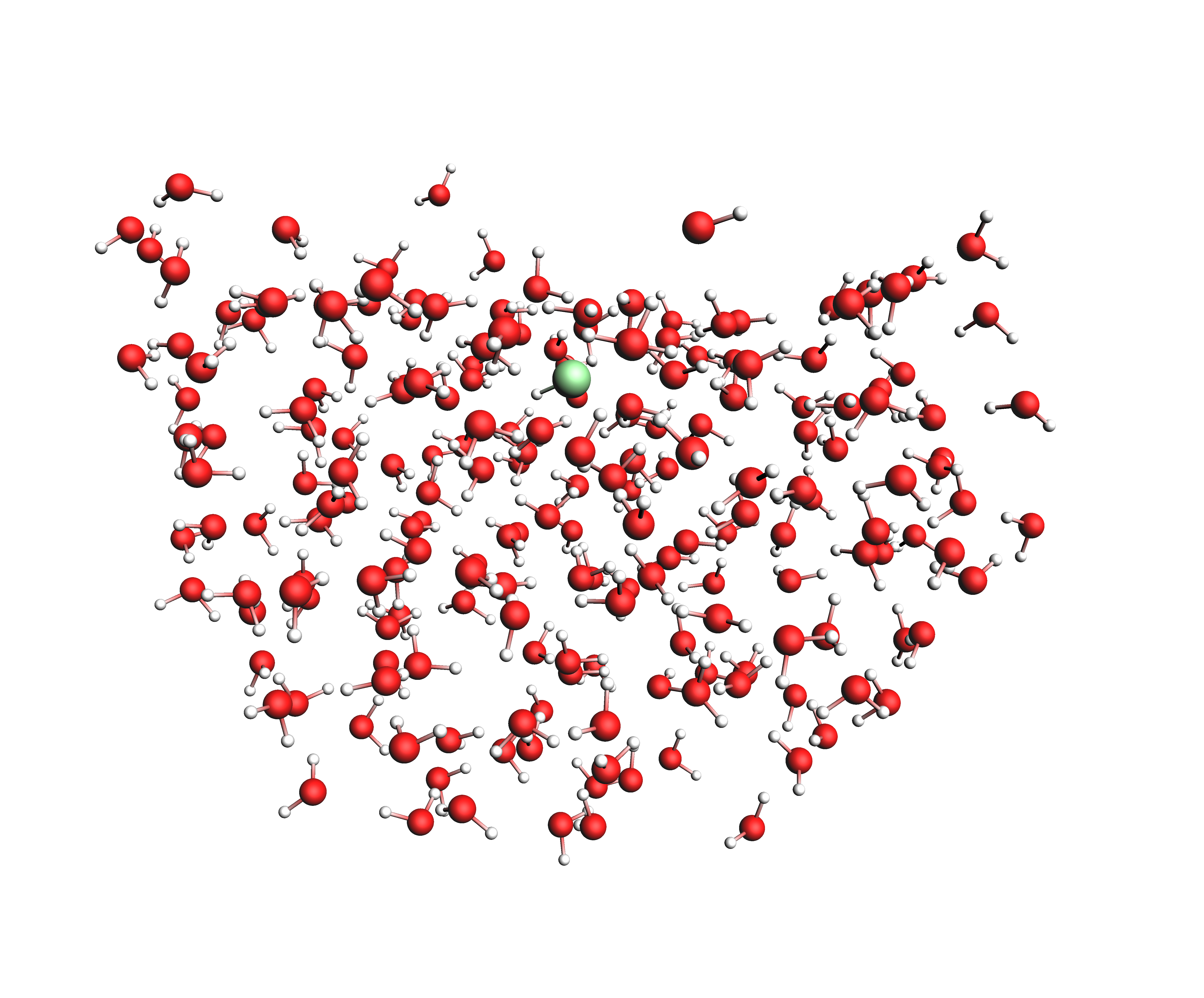}
\caption*{250 K}
\end{minipage}
\caption{\label{fig:temperature-hcl-structures} Structures for HCl from a snapshot from CMD simulations of HCl on ice at 235K (top) and 250K (bottom). The corresponding CC-in-DFT core electron binding energies for each configuration are given in table~\ref{tab:binding-energies-higher-temp-snaps}}.
\end{figure*}

From these simulations, we have extracted one snapshot for each temperature, and proceeded to the optimization of the HCl position while constraining all the water systems to remain fixed. From the resulting structures (figure~\ref{fig:temperature-hcl-structures}), we carried out CC-in-DFT calculations to obtain the binding energies shown in table~\ref{tab:binding-energies-higher-temp-snaps}.

\begin{table*}
\caption{\label{tab:binding-energies-higher-temp-snaps} $^2$DCG$^M$ CC-in-DFT chlorine core binding energies (in eV) for the CMD snapshots shown in figures~\ref{fig:temperature-hcl-structures} and~\ref{fig:temperature-cl-structures}. Apart from the energies for the individual edges, we provide the core binding energy shift ($\Delta_{BE}$, in eV) between HCl and Cl$^{-}$.}
\begin{tabular}{llllll}
\hline
\hline
        &       & \multicolumn{4}{c}{Core Binding Energies} \\
\cline{3-5}
Species & T (K) &   K     &  L$_1$  &  L$_2$  &   L$_3$ \\
\hline
HCl     &  235  & 2833.35 &  280.12 &  209.45 &  207.82 \\
Cl$^{-}$&      & 2825.20 &  271.79 &  201.16 &  199.53 \\
$\Delta_{BE}$&       &    8.15 &    8.33 &    8.29 &    8.29 \\
&&&&&\\
HCl     &  250  & 2833.60 &  280.38 &  209.71 &  208.08 \\
Cl$^{-}$&      & 2825.26 &  271.88 &  201.25 &  199.62 \\
$\Delta_{BE}$&       &    8.34 &    8.50 &    8.46 &    8.46 \\
\hline
\hline
\end{tabular}
\end{table*}

\clearpage

\clearpage

\begin{figure*}[htp!]
\centering
\begin{minipage}{0.67\textwidth}
\includegraphics[width=\linewidth]{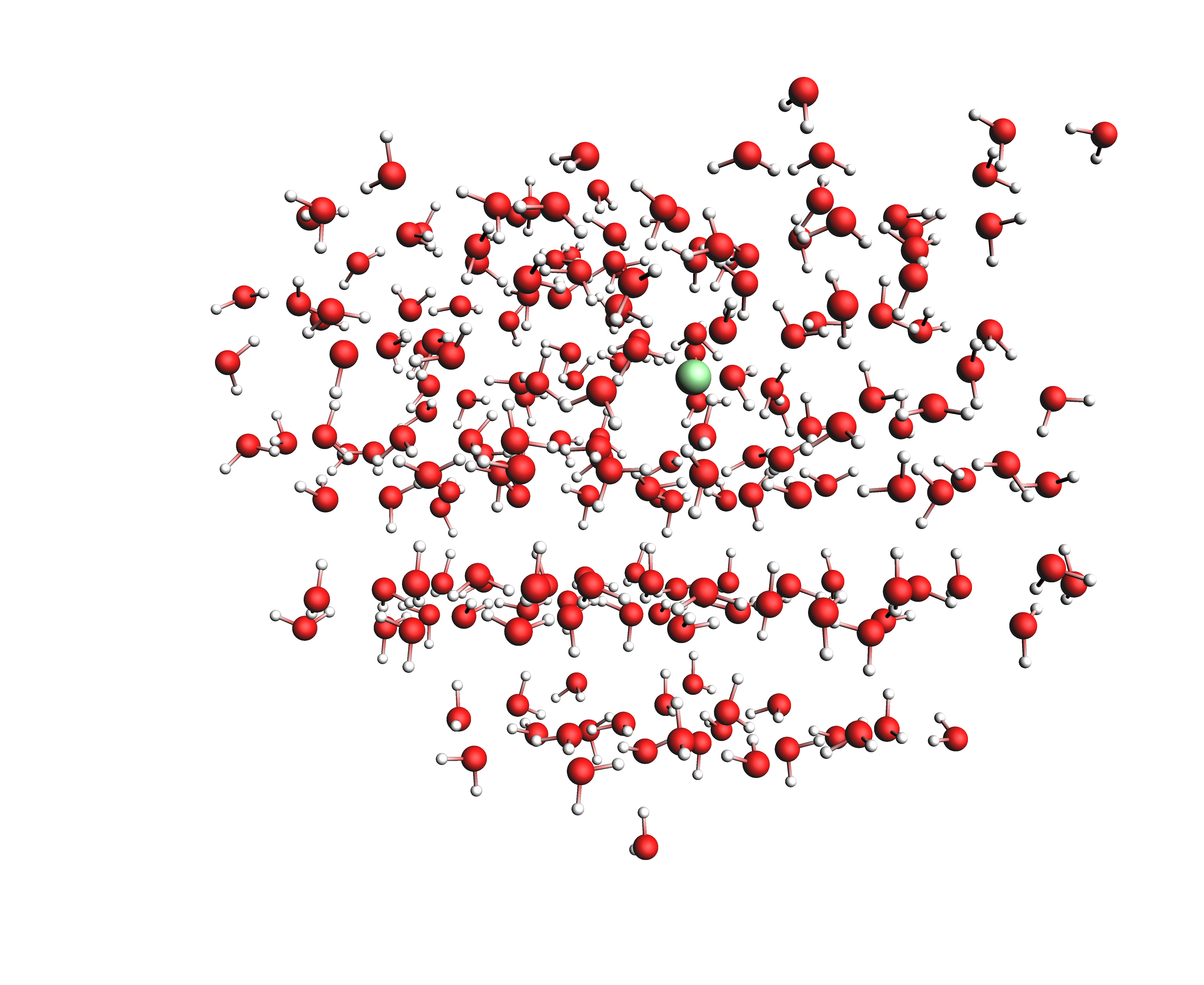}
\caption*{235 K}
\end{minipage}
\begin{minipage}{0.67\textwidth}
\includegraphics[width=\linewidth]{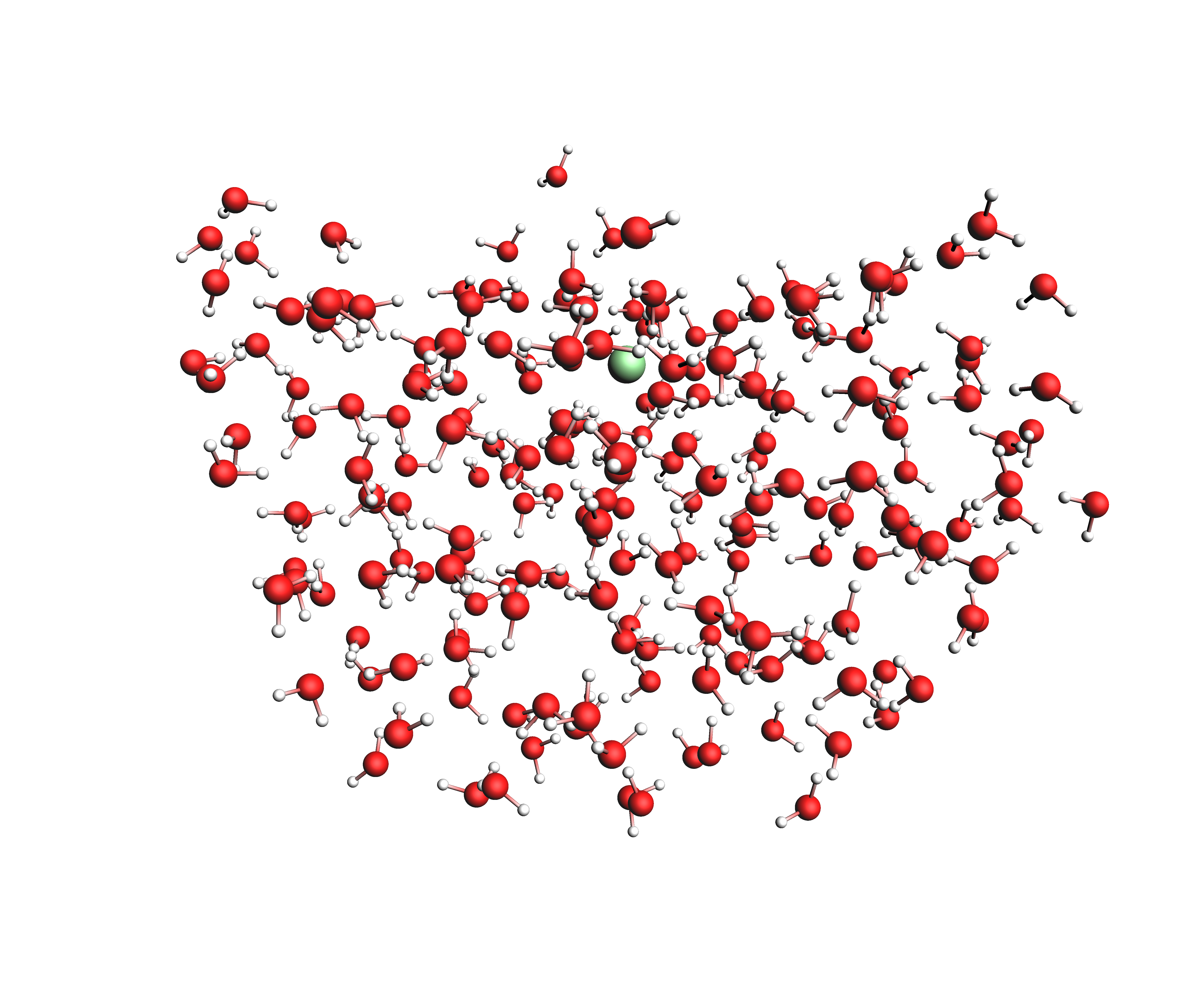}
\caption*{250 K}
\end{minipage}
\caption{\label{fig:temperature-cl-structures} Structures for Cl$^-$ from a snapshot from CMD simulations of HCl on ice at 235K (top) and 250K (bottom). The corresponding CC-in-DFT core electron binding energies for each configuration are given in table~\ref{tab:binding-energies-higher-temp-snaps}}.
\end{figure*}

\clearpage

\section{Effect of the counter-ion (H$^+$/H$_3$O$^+$) on Cl binding energies}

We investigated the effect of the counter-ion for Cl binding energies in HCl by constructing two models: (a) one in which we consider the Cl$^-$ ion in gas-phase, and mimic the effect of H$^+$ a point charge was placed at different distances from the anion. With this model we have aimed both to have an (electrostatic) embedding model, in line with the embedding approaches in the paper; and (b) a HCl water droplet model, based on a snapshot of the Cl$^-$ water droplet model, in which we started out by placing the H$^+$ species at different positions, near different heavy centers (Cl, O), and then proceeded with standard DFT (= without embedding) geometry optimization of the position of the added hydrogen while constraining either all other positions to remain the same (in the case of HCl), or optimizing the positions of the hydrogens attached to the oxygen center we attached the proton to. 

\begin{figure*}[htp!]
\centering
\begin{minipage}{0.45\textwidth}
\includegraphics[width=\linewidth]{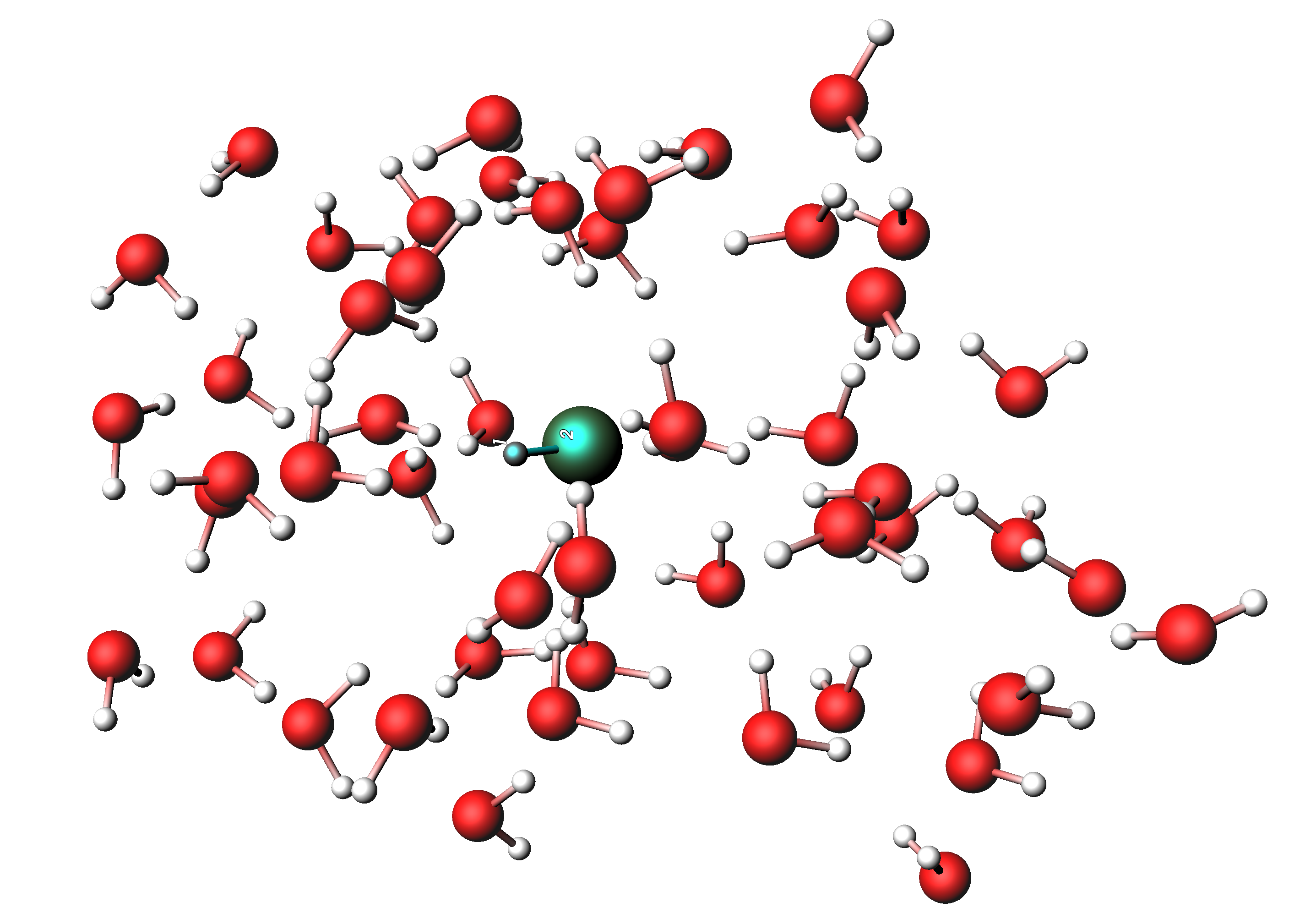}
\caption*{(a)}
\end{minipage}
\begin{minipage}{0.45\textwidth}
\includegraphics[width=\linewidth]{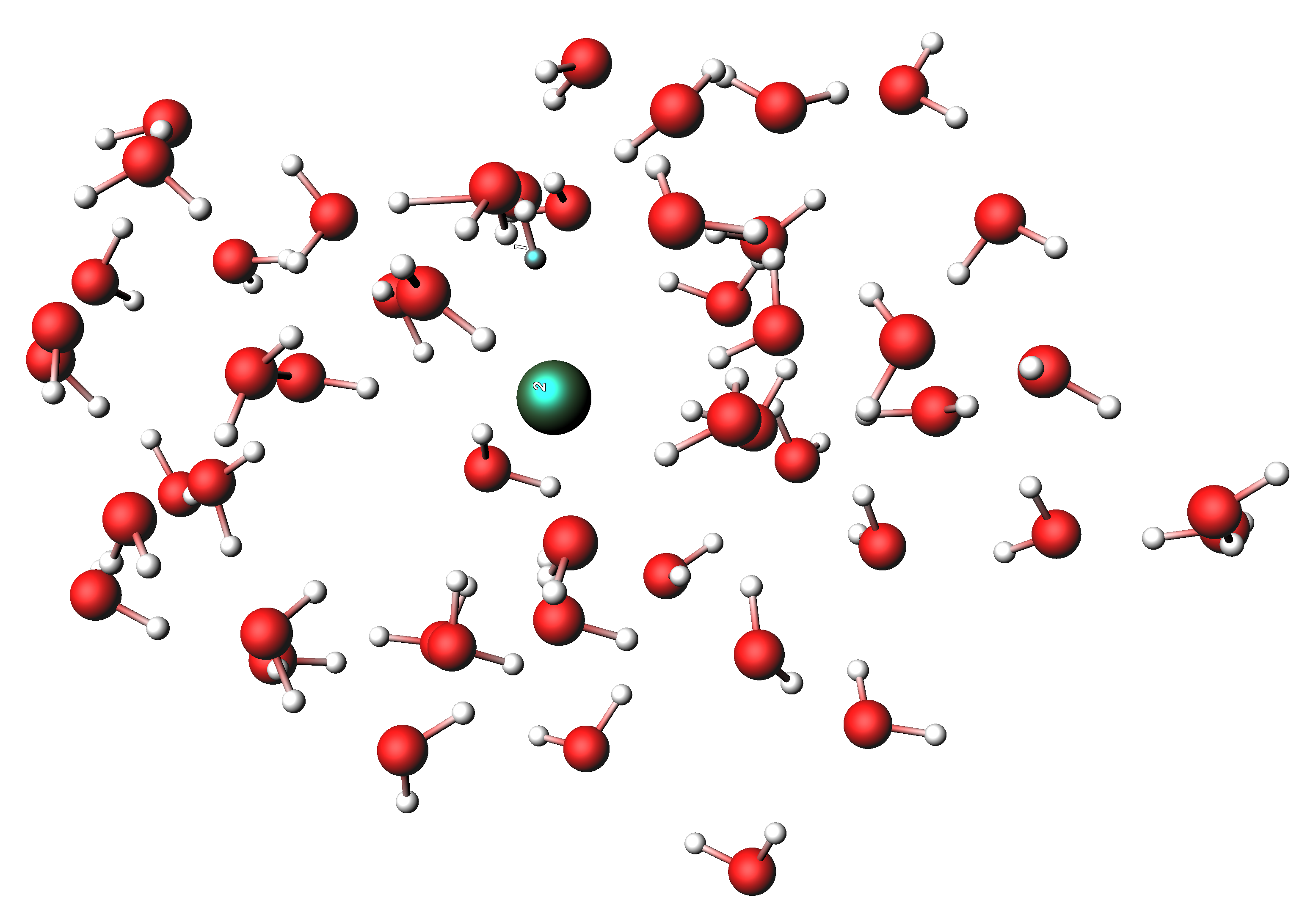}
\caption*{(b)}
\end{minipage}
\begin{minipage}{0.45\textwidth}
\includegraphics[width=\linewidth]{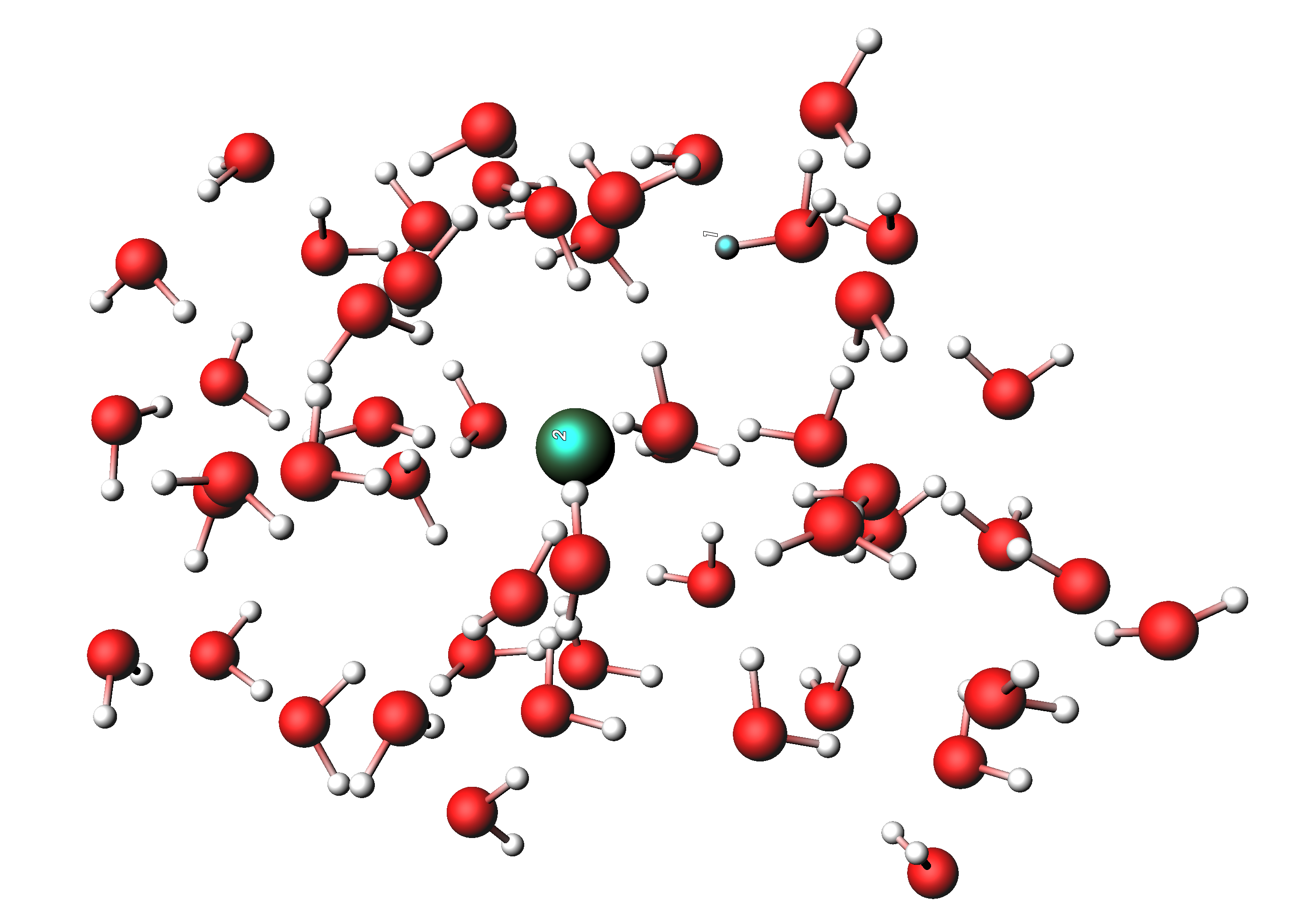}
\caption*{(c)}
\end{minipage}
\begin{minipage}{0.45\textwidth}
\includegraphics[width=\linewidth]{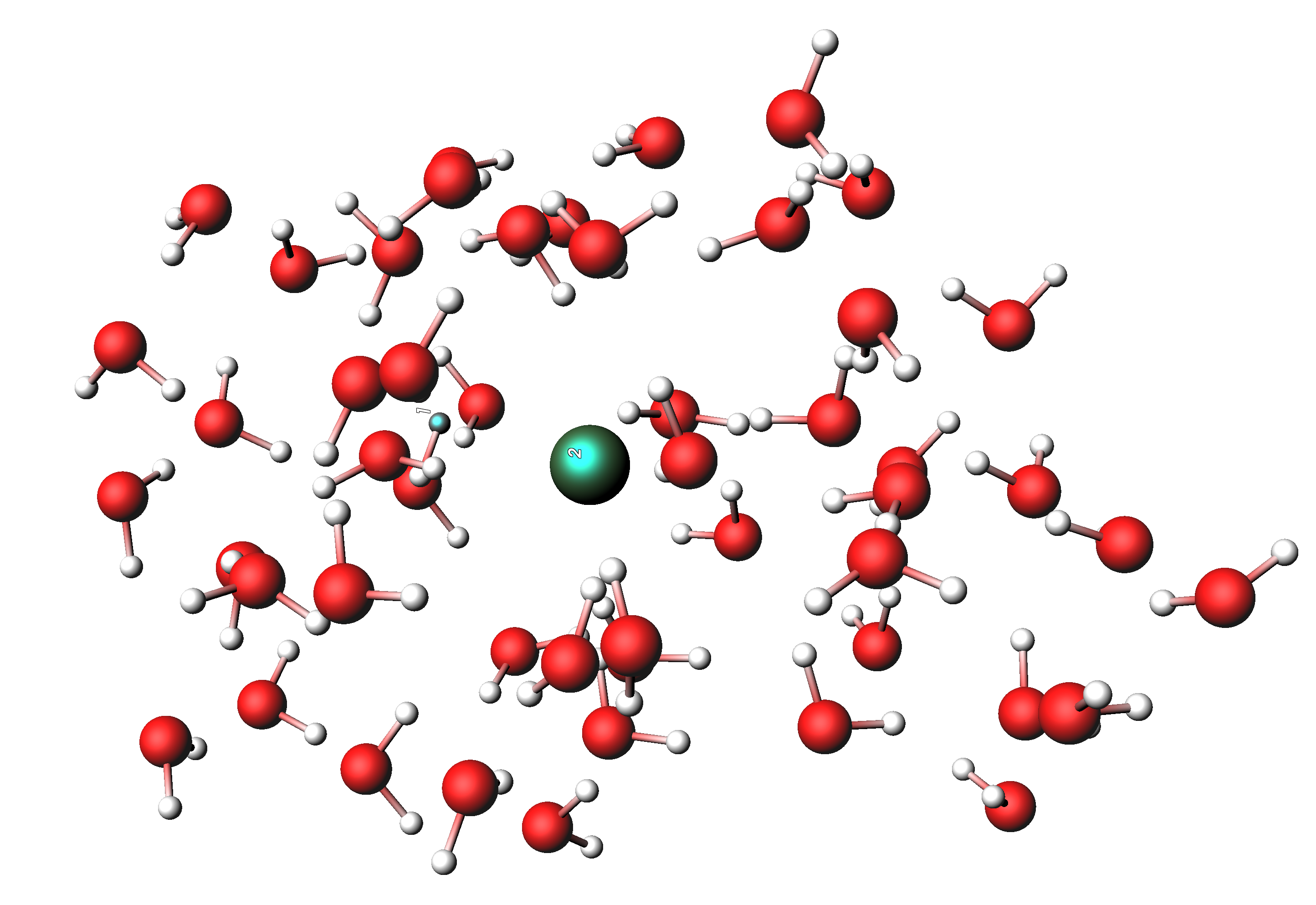}
\caption*{(d)}
\end{minipage}
\begin{minipage}{0.45\textwidth}
\includegraphics[width=\linewidth]{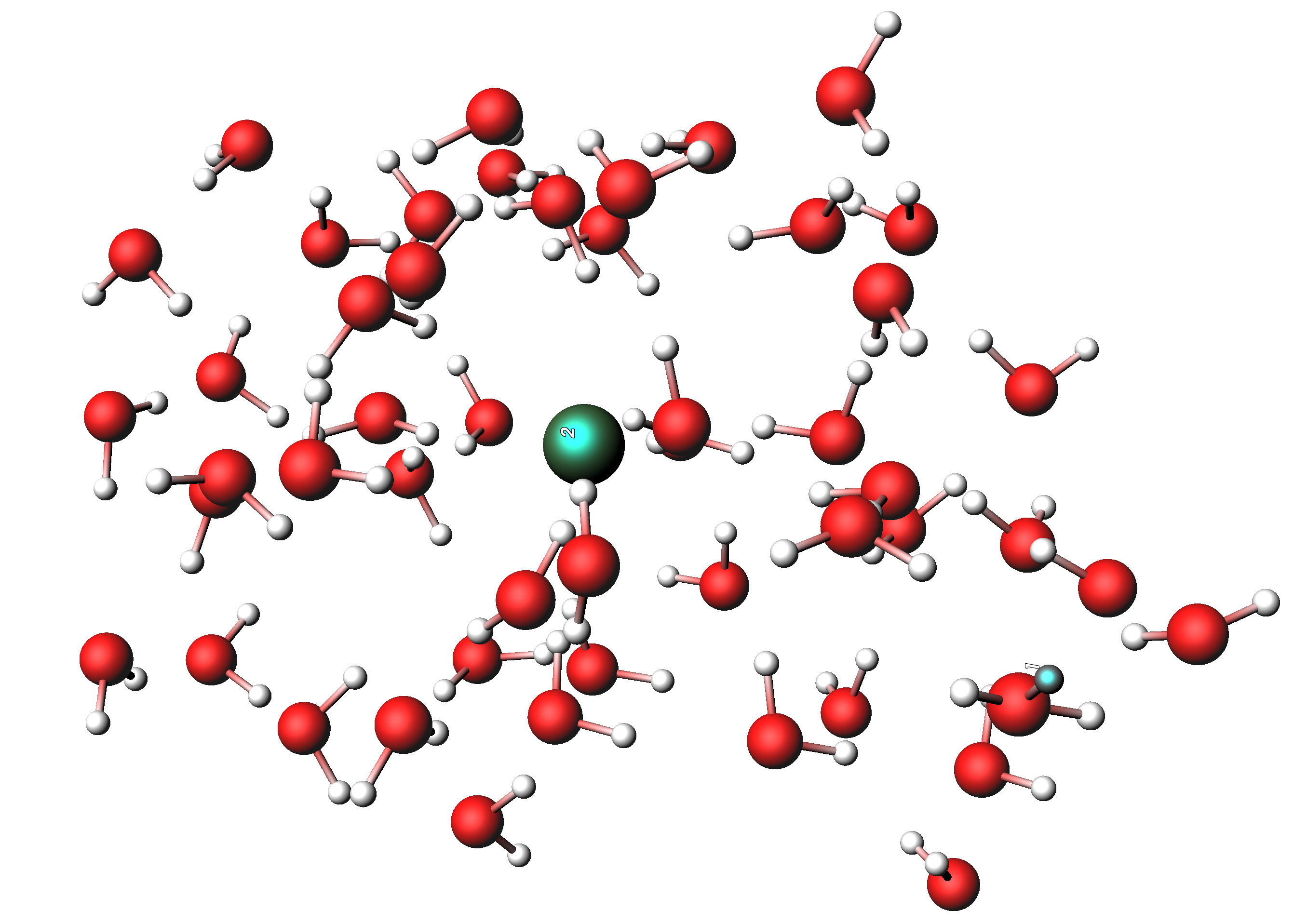}
\caption*{(e)}
\end{minipage}
\caption{\label{fig:hydronium-structures} Structures for [HCl(H$_2$O)$_{50}$] droplet models, derived from a parent [Cl(H$_2$O)$_{50}$]$^-$ droplet model structure. In these HCl droplet models, the hydrogen is found at different distances ($r$) to the central Cl$^-$: (a) $r = 1.306$\AA; (b) $r = 2.559$\AA; (c) $r = 3.484$\AA; (d) $r =5.690$\AA; and (e) $r = 10.256$\AA. In these figures, the chloride and corresponding hydrogen are highlighted.}
\end{figure*}

Due to the relatively long computation times for each constrained geometry optimization in model (b), we have only explored four configurations forming species resembling the H$_3$O$^+$ species, and one in which molecular HCl is formed (see figure~\ref{fig:hydronium-structures}), that would correspond to different H-Cl distances. 

We have chose to employ the droplet model since there is relatively little spread in the calculated core-binding energies with respect to the snapshot employed, and thus working with a single snapshot would be less prone to bias than working with the ice structures.

Apart from results shown in table~\ref{tab:binding-energies-hydronium-full}, we have explored other internuclear distances for the electrostatic model, to see how it compared to quantum mechanical calculations on HCl. 

At the gas-phase equilibrium structure ($r=1.27$\AA), the electrostatic model underestimates the (Koopmans theorem approximation to the) 1s binding energy with respect to the fully quantum mechanical model by about 1 eV (2748.9 eV for HCl and 2747.8 eV for the electrostatic embedding model), and at a slightly longer bond distance ($r = 1.4957$\AA) by about 1.5 eV (2749.1 eV for HCl and 2747.6 eV for the electrostatic embedding model). At intermediate distances of $r = 15$\AA and beyond, on the other hand, the agreement with isolated Cl$^-$ improves (overestimation of about 1 eV for $r = 15$\AA, 0.7 eV for $r = 20$\AA, 0.6 eV for $r = 25$\AA, 0.3 eV for $r = 50$\AA~and slightly over 0.1 eV for $r = 100$\AA.

\begin{table*}
\caption{\label{tab:binding-energies-hydronium-full} Approximate core electron binding energies (derived from supermolecular Scalar ZORA PBE orbital energies, in eV) for [Cl(H$_2$O)$_{50}$]$^-$ and [HCl(H$_2$O)$_{50}$] droplet model structure. For the latter, the binding energies are given as a function of the distances ($r$, in \AA) of the added hydrogen atom with respect to the chlorine atom. Given the lack of orbital relaxation due to the core hole formation in such calculations, we also present the core binding energy shift ($\Delta_{BE}$, in eV) between Cl$^{-}$ and the HCl systems, as this has been shown in the paper to be a robust way to estimate chemical shifts irrespective of the level of theory employed. To complement the droplet models, we also present results for a gas-phase model, in which a +1 point charge is added at the same distances $r$ from the chlorine atom.}
\begin{tabular}{l r rrr p{0.1cm} rrr}
\hline
\hline
&
& \multicolumn{7}{c}{Core Binding Energies} \\
\cline{3-9}
&
& \multicolumn{3}{c}{droplet model} && \multicolumn{3}{c}{gas-phase model} \\
\cline{3-5}
\cline{7-9}
System &   $r$     &    K       &       L$_1$   &       L$_2$/L$_3$ && K        &       L$_1$   &       L$_2$/L$_3$\\
\hline
HCl     &       1.306   &       2753.11 &       255.81  &       194.84  &&      2747.88 &       250.64  &       189.58  \\
        &       2.559   &       2748.68 &       251.33  &       190.20  &&      2744.99 &       247.62  &       186.53  \\
        &       3.489   &       2748.44 &       251.09  &       189.97  &&      2743.44 &       246.08  &       184.96  \\
        &       5.690   &       2747.95 &       250.60  &       189.48  &&      2741.81 &       244.45  &       183.31  \\
        &       10.256  &       2747.23 &       249.92  &       188.79  &&      2740.69 &       243.32  &       182.18  \\
        &               &               &               &               &&              &               &               \\
Cl$^-$  &       $\infty$&       2745.63 &       248.28  &       187.15  &&      2739.29 &       241.91  &       180.78  \\
        &               &               &               &               &&              &               &               \\
$\Delta_{BE}$&  1.306   &       7.48    &       7.53    &       7.69    &&      8.59    &       8.73    &       8.80    \\
        &       2.559   &       3.05    &       3.06    &       3.05    &&      5.70    &       5.71    &       5.75    \\
        &       3.489   &       2.81    &       2.81    &       2.82    &&      4.16    &       4.17    &       4.18    \\
        &       5.690   &       2.32    &       2.33    &       2.33    &&      2.53    &       2.53    &       2.54    \\
        &       10.256  &       1.60    &       1.64    &       1.64    &&      1.40    &       1.40    &       1.40    \\
\hline
\hline
\end{tabular}
\end{table*}

\clearpage

